\documentclass{aa}  

\usepackage{graphicx}
\usepackage{txfonts}
\usepackage[breaklinks=true]{hyperref} 
\hypersetup{
  colorlinks   = true, %Colours links instead of ugly boxes
  urlcolor     = red, %Colour for external hyperlinks
  linkcolor    = blue, %Colour of internal links
  citecolor   = blue %Colour of citations
}
\newcommand{\lyaw}{\hbox{Ly$\alpha$}\,$\lambda1216$}
\newcommand{\lya}{\hbox{Ly$\alpha$}}
\newcommand{\haw}{\hbox{H$\alpha$}\,$\lambda6563$}
\newcommand{\hbw}{\hbox{H$\beta$}\,$\lambda4861$}
\newcommand{\oiid}{[\ion{O}{ii}]$\lambda\lambda$3726,3729}
\newcommand{\oiiibpt}{[\ion{O}{iii}]$\lambda5007$}

\newcommand{\oiiisf}{\ion{O}{iii}]}
\newcommand{\oiiib}{\ion{O}{iii}]$\lambda1661$}
\newcommand{\oiiir}{\ion{O}{iii}]$\lambda1666$}
\newcommand{\oiiid}{\ion{O}{iii}]$\lambda\lambda$1661,1666}
\newcommand{\nv}{\ion{N}{v}}
\newcommand{\nvl}{\ion{N}{v}$\lambda1240$}
\newcommand{\civd}{\ion{C}{iv}$\lambda\lambda$1548,1551}
\newcommand{\civt}{\ion{C}{iv}$\lambda1550$}
\newcommand{\civ}{\ion{C}{iv}}
\newcommand{\ciia}{\ion{C}{ii}$\lambda1335$}
\newcommand{\ciiib}{[\ion{C}{iii}]$\lambda1907$}
\newcommand{\ciiir}{\ion{C}{iii}]$\lambda1909$}
\newcommand{\ciiid}{[\ion{C}{iii}]$\lambda1907$+\ion{C}{iii}]$\lambda1909$}

\newcommand{\ciii}{\ion{C}{iii}]}
\newcommand{\cii}{\ion{C}{ii}}
\newcommand{\heiil}{\ion{He}{ii}$\lambda1640$}
\newcommand{\heii}{\ion{He}{ii}}
\newcommand{\oisiliit}{\ion{O}{i}$\lambda1302$ +\ion{Si}{ii}$\lambda1304$}
\newcommand{\oisilii}{\ion{O}{i} + \ion{Si}{ii}}
\newcommand{\siliia}{\ion{Si}{ii}$\lambda1260$}
\newcommand{\siliib}{\ion{Si}{ii}$\lambda1527$}
\newcommand{\siliic}{\ion{Si}{ii*}$\lambda1533$}
\newcommand{\silii}{\ion{Si}{ii}}
\newcommand{\siliis}{\ion{Si}{ii*}}
\newcommand{\siliii}{\ion{Si}{iii}]}
\newcommand{\siliiib}{[\ion{Si}{iii}]$\lambda1883$}
\newcommand{\siliiir}{\ion{Si}{iii}]$\lambda1892$}
\newcommand{\siliiid}{[\ion{Si}{iii}]$\lambda1883$+\ion{Si}{iii}]$\lambda1892$}

\newcommand{\mgii}{\ion{Mg}{ii}}

\newcommand{\silivl}{\ion{Si}{iv}$\lambda1403$}
\newcommand{\siliv}{\ion{Si}{iv}}

\newcommand{\magphys}{\texttt{MAGPHYS}}
\newcommand{\platefit}{\texttt{PLATEFIT}}
\newcommand{\udften}{\textsf{udf-10}}
\newcommand{\mosaic}{\textsf{mosaic}}

\usepackage[normalem]{ulem}

\begin{document}

   \title{The MUSE Hubble Ultra Deep Field Survey}

   \subtitle{XV. The mean rest-UV spectra of \lya\ emitters at z>3 \thanks{Based on observations made with ESO telescopes at the La Silla
Paranal Observatory under programs 094.A-0289(B), 095.A-0010(A),
096.A-0045(A) and 096.A-0045(B).}}

\titlerunning{The mean rest-UV spectra of MUSE HUDF LAEs at $3 < z < 4.6$}

   \author{Anna Feltre\inst{1,2}
   \and
   Michael V. Maseda\inst{3}
   \and
   Roland Bacon\inst{4}
   \and
   Jayadev Pradeep\inst{5}
   \and
   Floriane Leclercq\inst{6}
   \and
   Haruka Kusakabe\inst{6}
   \and
   Lutz Wisotzki\inst{7}
   \and 
   Takuya Hashimoto\inst{8,4}
   \and
   Kasper B. Schmidt\inst{7}
   \and
   Jeremy Blaizot\inst{4}
   \and
   Jarle Brinchmann\inst{9,3}
   \and 
   Leindert Boogaard\inst{3}
   \and
   Sebastiano Cantalupo\inst{10}
   \and
   David Carton\inst{4}
   \and
   Hanae Inami\inst{11}
   \and
   Wolfram Kollatschny\inst{12}
   \and
   Raffaella A. Marino\inst{10}
   \and
   Jorryt Matthee\inst{10}
   \and
   Themiya Nanayakkara\inst{3,13}
   \and
   Johan Richard\inst{4}
   \and
   Joop Schaye\inst{3}
   \and
   Laurence Tresse \inst{4}
   \and 
   Tanya Urrutia\inst{7}
    \and
   Anne Verhamme\inst{6}
   \and
   Peter M. Weilbacher\inst{7}
  }

   \institute{SISSA, Via Bonomea 265, 34136 Trieste, Italy  
   \and
   INAF - Osservatorio di Astrofisica e Scienza dello Spazio di Bologna, Via P. Gobetti 93/3, 40129 Bologna, Italy \email{anna.feltre@inaf.it}
   \and
   Leiden Observatory, Leiden University, P.O. Box 9513, 2300 RA Leiden, The Netherlands
   \and
   Univ Lyon, Univ Lyon1, Ens de Lyon, CNRS, Centre de Recherche Astrophysique de Lyon UMR5574, F-69230, Saint-Genis-Laval, France
     \and
     Department of Earth and Space Sciences, Indian Institute of Space Science \& Technology, Thiruvananthapuram 695547, India
     \and
   Observatoire de Gen\`eve, Universite de Gen\`eve, 51 Ch. des Maillettes, 1290 Versoix, Switzerland
    \and
    Leibniz-Institut f\"ur Astrophysik Potsdam (AIP), An der Sternwarte 16, 14482 Potsdam, Germany
    \and 
    Tomonaga Center for the History of the Universe (TCHoU), Faculty of Pure and Applied Sciences, University of Tsukuba,Tsukuba, Ibaraki 305-8571, Japan 
    \and
    Instituto de Astrof\'isica e C\^iencias do Espa\c{c}o, Universidade  do  Porto,  CAUP,  Rua  das  Estrelas,  PT4150-762Porto, Portugal
    \and
    Department of Physics, ETH Z\"urich, Wolfgang-Pauli-Strasse 27, 8093 Z\"urich, Switzerland
    \and
    Hiroshima Astrophysical Science Center, Hiroshima University, 1-3-1 Kagamiyama, Higashi-Hiroshima, Hiroshima 739-8526, Japan
    \and
    Institut f\"ur Astrophysik, Universit\"at G\"ottingen,Friedrich-Hund Platz 1, D-37077 G\"ottingen, Germany
    \and Centre for Astrophysics and Supercomputing, Swinburne University of Technology, Hawthorn, Victoria 3122, Australia}
   \date{Received Month X, XXXX; accepted Month X, XXXX}

  \abstract{We investigated the ultraviolet (UV) spectral properties of faint Lyman-$\alpha$ emitters (LAEs) in the redshift range $2.9 \leq z \leq 4.6,$ and we provide material to prepare future observations of the faint Universe. We used data from the MUSE {\itshape Hubble} Ultra Deep Survey to construct mean rest-frame spectra of continuum-faint (median M$_{\rm UV}$ of $-$18 and down to M$_{\rm UV}$ of $-$16), low stellar mass (median value of $10^{8.4}$ M$_{\odot}$ and down to $10^{7}$ M$_{\odot}$) LAEs at redshift $z \gtrsim 3$. We computed various averaged spectra of LAEs, subsampled on the basis of their observational (e.g., \lya\ strength, UV magnitude and spectral slope) and physical (e.g., stellar mass and star-formation rate) properties. We searched for UV spectral features other than \lya, such as higher ionization nebular emission lines and absorption features. We successfully observed the \oiiir\ and \ciiid\ collisionally excited emission lines and the \heiil\ recombination feature, as well as the resonant \civd\ doublet either in emission or P-Cygni. We compared the observed spectral properties of the different mean spectra and find the emission lines to vary with the observational and physical properties of the LAEs. In particular, the mean spectra of LAEs with larger \lya\ equivalent widths, fainter UV magnitudes, bluer UV spectral slopes, and lower stellar masses show the strongest nebular emission. The line ratios of these lines are similar to those measured in the spectra of local metal-poor galaxies, while their equivalent widths are weaker compared to the handful of extreme values detected in individual spectra of $z>2$ galaxies. This suggests that weak UV features are likely ubiquitous in high $z$, low-mass, and faint LAEs. We publicly released the stacked spectra, as they can serve as empirical templates for the design of future observations, such as those with the James Webb Space Telescope and the Extremely Large Telescope.}
 

   \maketitle

\section{Introduction}

Ultraviolet (UV) line-emitting galaxies (at low and high redshift) are receiving a great deal of attention, as they are possible analogs of the faint, low-mass but numerous, distant star-forming galaxies, which are considered the best candidates to provide the hydrogren-ionizing budget to sustain the cosmic reionization down to $z\approx6$ \citep[e.g.,][]{Robertson2015, Bouwens2016, Finkelstein2019}. 
The rest-frame UV ($\lambda < 3200 \AA$) is rich in high-ionization metal lines, for example, \ciiid, \civt,\ and \nvl, which provide additional information to the rest-optical lines commonly explored to understand the nature of ionizing sources \citep[e.g.,][]{Steidel2016, Senchyna2017, Topping2019}. 
Spectral features, in absorption or in emission, provide valuable clues concerning the properties of the stellar populations of galaxies and the physical conditions in their interstellar medium (ISM).
While the exploitation of strong rest-optical emission lines for statistical studies of the ionized properties of galaxies is limited to $z \leq 3$, current optical and near-infrared ground-based spectrographs (e.g., the Multi Unit Spectroscopic Explorer, MUSE, and XShooter on the ESO-Very Large Telescope and MOSFIRE of the Keck Observatory) are now providing high-quality rest-UV spectra for targeted samples of galaxies at earlier epochs (from $z\approx3$ out to $z\approx6-7$).

A fast-growing number of observational studies of $z\gtrsim2$ galaxies present spectra showing strong UV emission features, in addition to the \lya\ line. While the detection of multiple UV emission lines on single spectra mainly relies on measurements performed on gravitationally lensed sources \citep[e.g.,][]{Stark2015b, Berg2018}, large deep surveys, such as VUDS \citep{LeFevre2015} and VANDELS \citep{McLure2018,Pentericci2018}, have just enabled the assembly of non-lensed samples of UV line-emitting galaxies at $z\approx2-4$ \citep[e.g.,][]{Amorin2017, Nakajima2018a, LeFevre2019, Marchi2019}. High-ionization UV nebular emission lines (e.g., \civt\ and \heiil) have been identified in the spectra of young galaxies at high redshift, from $z\approx2$ up to $z\approx7$ \citep[e.g.,][]{Stark2014, Stark2015b, Mainali2017, Berg2018, Nanayakkara2019, Vanzella2010, Vanzella2016, Vanzella2017, Vanzella2020}, highlighting the central role of young ($\leq$10 Myr) and massive stars in contributing to the spectral energy distribution (SED) of high-$z$ galaxies, and triggering  interest in the ionizing properties of these extreme line emitters. 
These observations challenge current stellar population synthesis models, as they do not predict hard-enough ionizing radiation to account for the strong \heii\ emission observed in local metal-poor galaxies and  $z>2$ line-emitting galaxies \citep[e.g.,][]{Senchyna2017, Nanayakkara2019, Plat2019}.

Obtaining individual high signal-to-noise ratio (S/N) detections of multiple UV lines (other than \lya) at high $z$ is extremely challenging with current instrumentation, and a standard procedure for studying spectral features too faint to be detected in single spectra is spectral stacking, which provides higher S/N data by co-adding individual lower S/N spectra \citep[e.g.,][]{Shapley2003, Steidel2010, Steidel2016, Berry2012, Jones2012, Rigby2018, Nakajima2018b, Pahl2020, Khusanova2019, Thomas2019, Trainor2019}. 
By splitting $\sim$1000 Lyman-break galaxies (LBGs) into subsamples based on \lyaw\ (hereafter \lya) equivalent width (EW), UV spectral slope, and UV magnitude, \cite{Shapley2003} studied how the spectral properties of the composite spectra of each subsample depend on other galaxy properties. \cite{Berry2012} and \cite{Jones2012} performed similar analyses on $\approx80$ UV-bright star-forming galaxies at $2.0<z<3.5$ and on $\approx 80$ LBGs in the range $3<z<7$, respectively. One of the main outcomes from these works is that galaxies with stronger \lya\ emission show bluer UV continuum slopes, lower stellar masses, and spectra with weaker low-ionization absorption-line profiles and more prominent nebular emission lines. More recently, \cite{Pahl2020} computed stacked spectra by grouping a sample of 375 star-forming galaxies at $4<z<5.5$ in different bins of \lya\ EW and found a decrease in the low-ionization absorption-line strength with increasing EW of \lya\ emission up to $z\approx5$, confirming previous trends. 
\cite{Rigby2018} publicly released the composite spectra of 14 highly magnified star-forming galaxies at $1.6<z<2.3$. They are extremely rich in absorption and nebular emission features and are to be used as diagnostics of the physical properties of the stellar population, the physical conditions of ionized gas, and outflowing winds within these galaxies. \cite{Nakajima2018b} studied the nebular emission features of  $\sim100$  Lyman-alpha emitters (LAEs) at $z\approx3.1$, confirming earlier suggestions of a correlation between the strength of \lya\ and \ciii\ emission \citep[][]{Shapley2003, Stark2014} and finding similar trends with the UV luminosities and colors. The results from \cite{Nakajima2018b} support earlier suggestions that LAEs have a significantly larger ionizing production efficiency ($\xi_{\rm ion}$) than LBGs \citep[see also][]{Lam2019}. Recently, \cite{Saxena2019} presented the stacked spectra of \heii\ emitters at $z \approx 2.5 - 5.0$, identified in the deep VANDELS ESO public spectroscopic survey \citep[][]{McLure2018, Pentericci2018}, finding no significant difference between galaxies with and without \heii\ emission in terms of  stellar masses, star formation rates (SFRs) and rest-frame UV magnitudes.

All of these studies rely implicitly on pre-selections that may not result in a sample that is representative of the overall population.  The Multi-Unit Spectroscopic Explorer \cite[MUSE;][]{Bacon2010} on the Very Large Telescope (VLT), however, has recently enabled the detection and analysis of un-targeted samples of LAEs at $2.9 < z < 6.7$ \cite[e.g.,][]{Bacon2015, Wisotzki2016, Leclercq2017, Hashimoto2017b, Drake2017,  Maseda2018, Marino2018, Maseda2020}, that is, selecting galaxies only on their \lya\ emission line. In particular, the MUSE {\it Hubble} Ultra Deep Field (HUDF) Survey \citep{Bacon2017} uncovers an unprecedented number of the UV faint (M$_{\rm UV}$  at 1500 \AA\ down to $-$16), low stellar mass (down to $\sim 10^{7} \, $M$_{\odot}$) LAEs at $z>3$, reaching UV luminosities (and stellar masses) two orders of magnitude fainter (smaller) than previous studies. 
A systematic analysis of how the UV spectral features of these faint and low-mass LAEs depends on the observed and physical properties of galaxies and how they compare with those observed in brighter and more massive sources is missing from the literature. 

In this work, we present unweighted mean spectra of the MUSE HUDF LAEs grouped in subsamples on the basis of observational (e.g., \lya\ strengths and EW, UV magnitude, and UV spectral slopes) and physical (e.g., stellar mass, SFR) galaxy properties. The aim of this paper is to study variations of the spectral properties among the different LAE subsamples, with a particular focus on nebular line emission. The stacked spectra are publicly available  together with this paper and can serve as empirical templates for the design of future observations (e.g., those with the JWST and the ELT). 

The spectroscopic dataset is described in Sect. \ref{sec:data}, and the stacking procedure in Sect. \ref{sec:stacking}. Section \ref{sec:results} illustrates the different mean spectra and their spectral properties, with particular focus on nebular emission lines such as \heiil, \oiiid, \siliiid, and \ciiid,\ and the \civd\ resonant doublet.  The dependencies of these features on the galaxy properties, and a comparison with the literature are discussed in Sect. \ref{sec:discussion}, followed  by the conclusions in Sect. \ref{sec:conclusions}. Throughout the paper, we use the AB flux normalization \citep{Oke1983}, and we adopt the cosmological  parameters from \cite{Planck2016}, ($\Omega_{\rm M}$, $\Omega_{\Lambda}$, $H_{\rm 0}$) = (0.308, 0.692, 67.81).

\section{Dataset}\label{sec:data}

\subsection{MUSE HUDF Survey and sample selection}\label{sec:sample}

We made use of spectroscopic observations from the MUSE HUDF Survey \citep[][]{Bacon2017}, which consists of a mosaic of nine pointings ($\sim 3' \times 3'$) with 10-hour exposure (\mosaic) in the HUDF and an additional single deeper exposure of 31 hours (\udften) within the same region. We opted for an upgraded version of the \mosaic\ and \udften\ datacubes (version 1.0), which, compared to version 0.42 used in \citet[]{Bacon2017}, is created with an enhanced data reduction process resulting in an improved sky subtraction and S/N of the extracted spectra (Bacon et al., in prep).

We used the redshift measurements from the first MUSE HUDF Survey Data Release (DR1) catalog \citep[][hereafter I17]{Inami2017}, which combines galaxies from the \mosaic\ and \udften, avoiding duplicates. 
We selected LAEs in the redshift range $2.9 \leq z \leq 4.6$ to include the following UV lines in the MUSE spectral coverage: \nvl, \civd, \heiil, \oiiid, \siliiid,\ and \ciiid\ (hereafter \nv, \civ, \heii, \oiiisf, \siliii\ and \ciii, respectively). Throughout the paper, we explicitly report the rest wavelength of a given transition when referring to a specific component of the line doublets.

We assembled our sample of LAEs from the MUSE HUDF Survey according to the following selection procedure:

\begin{itemize}
    \item We first selected 488 MUSE HUDF targets with secure redshift measurements (\texttt{CONFID=2 and 3} as explained in I17) in the $2.9 \leq z \leq 4.6$ range and defined as LAEs (\texttt{TYPE=6} in I17).
    \item We excluded a total of 88 sources with multiple HST ({\it Hubble Space Telescope}) associations (i.e., blended sources) or without a HST counterpart \cite[cf.][]{Maseda2018}. The HST detections are necessary for inferring physical properties, such as stellar mass and SFR, from SED fitting to the broad-band photometry. 
    \item We removed two sources (ID 1056 and 6672 in the DR1 catalog of I17) that are defined as active galactic nuclei (AGN) on the basis of their X-ray emission from the 7 Ms Source Catalogs of the Chandra Deep Field-South Survey \citep[][see their Section 4.5 for source classification]{Luo2017}. We did not find any other evidence, either from MUSE spectra or from X-ray data, for strong AGN in this sample, as discussed in Sect. \ref{sec:agn_cont}.
    \item Finally, we required a minimal S/N of 5 for the  flux of the \lya\ line computed with the curve of growth (CoG) method \citep[e.g., Sect. 5.3.2 of][hereafer L17, see also Sect. \ref{sec:extractions} of this work]{Leclercq2017}, excluding 178 targets.
\end{itemize}

With this final step, we were left with 220 LAEs in the redshift range $2.9 \leq z \leq 4.6$ (62 and 158 in the \udften\ and \mosaic, respectively). 
Since we performed \lya\ flux measurements (Sect. \ref{sec:extractions}) on the new version (1.0) of the datacube, we compared our measurements with the \lya\ fluxes computed by L17 on version 0.42, both obtained with the CoG method, finding agreement within the 1$\sigma$ errors.
Our sample contains a higher number of LAEs in the redshift range of interest than the L17 catalog because of the minimal S/N of 5 imposed on the \lya\ flux, compared to the slightly more conservative value of 6 imposed by L17. We relaxed the S/N cut from 6 to 5 to reduce the bias toward brighter halos (see Fig. 1 of L17).  
The different number of sources between our LAEs and those of \cite{Hashimoto2017b}, hereafter H17, is due to i)  the additional requirement on the HST band detections imposed during the selection process (Sect. 2.3 and Table 1 of H17), and ii) their pre-selection of LAEs imposing an S/N cut of 5 on \lya\ fluxes measured with the \platefit\ tool \citep{Brinchmann2004, Tremonti2004} on the one-dimensional spectral extraction rather than applying the CoG method.

\subsection{Properties of LAEs}\label{sec:properties}

\begin{figure}[!h]
  \centering
   \includegraphics[width=8.8cm]{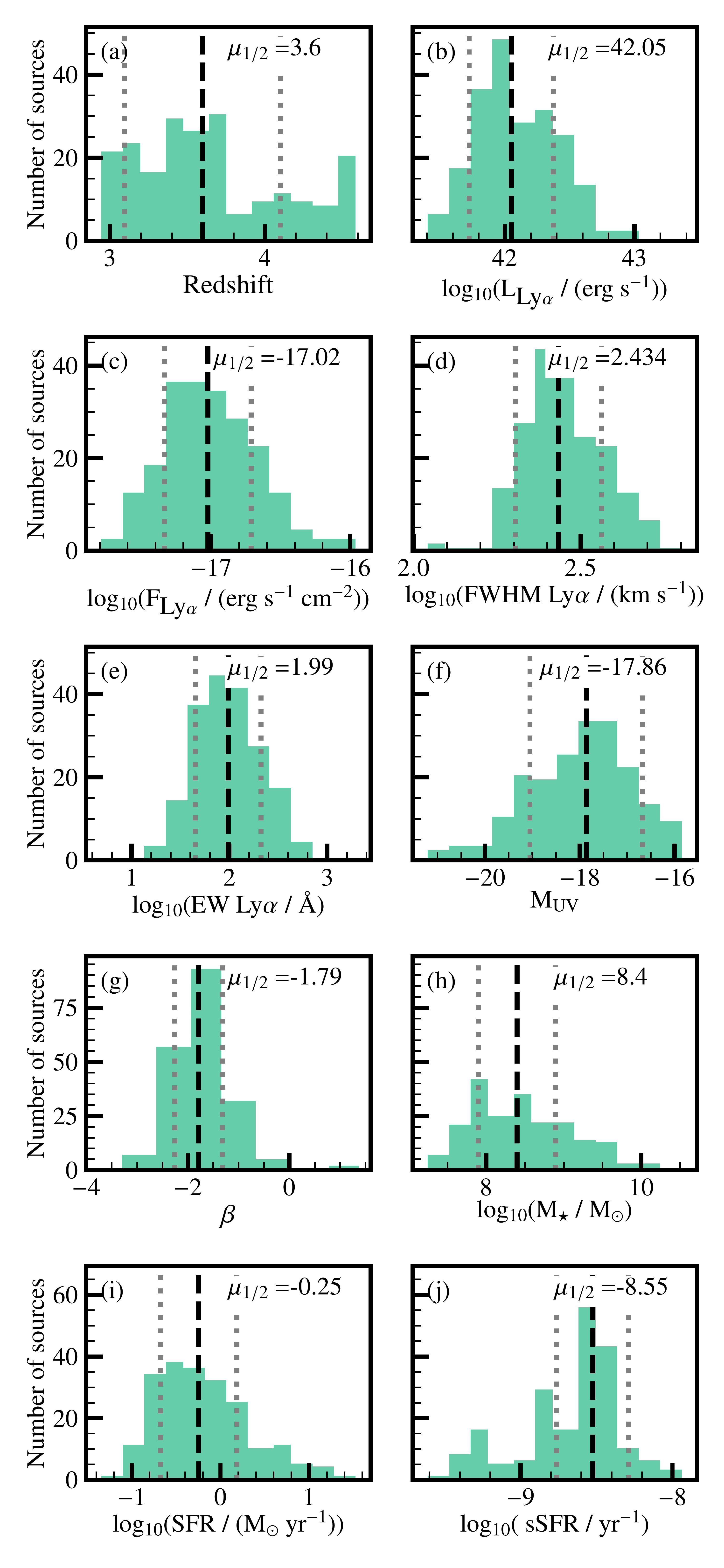}
   \caption{Distributions of observed and physical properties of the LAE sample described in Sect.~\ref{sec:sample}. From panel a) to j): redshift, \lya\ luminosity, \lya\ flux, \lya\ FWHM, rest-frame \lya\ EW, absolute UV magnitude, UV spectral slope, stellar mass, SFR, and sSFR. The \lya\ emission shown here has not been corrected by attenuation from dust. The vertical dashed black line indicates the median value, $\mu_{1/2}$ (reported in the top right), of the distributions used to select the subsamples of Sect.~\ref{sec:subsamples}. The vertical dotted gray lines indicate the median absolute deviation. }
            \label{Fig1}%
   \end{figure}

The redshift distribution of our LAEs is shown in Fig.~\ref{Fig1}, panel (a), along with the distributions of other observational and physical properties of our LAEs. The \lya\ luminosities and fluxes (panels (b) and (c), respectively) were computed by applying a CoG method on the new version (1.0) of the datacube. For the 198 LAEs showing a single-peaked \lya\ profile, we computed the full width at half maximum, FWHM, shown in Panel (d). The \lya\ EW, panel (e), is computed from the \lya\ fluxes and the continuum at 1216 \AA\ estimated from the UV magnitude at 1500 \AA\ and the $\beta$ slope \citep[Sect. 6.1 of H17, see also Sect. 2.2.3 of][]{Kusakabe2020}. 

The information on the absolute UV magnitude at 1500\AA, M$_{\rm UV}$, and the UV spectral slope $\beta$, panels (f) and (g), computed in H17 (see their Sect. 3.1) is available only for 178 objects in common with their sample. Using the same method as H17, we computed M$_{\rm UV}$ and $\beta$ for other 26 objects that had at least two HST detections above 2$\sigma$ in the HST passband filters listed in Table 2 of H17. 
We caution the reader, however, that the determination of the $\beta$ slope can be subject to high uncertainties, in particular for weaker sources, with errors as large as one (see Table 3 of H17).

The stellar mass, SFR, and specific SFR (sSFR) (panels (h), (i), and (j), respectively) were inferred via SED fitting using the high-$z$ extension of the code \magphys\ \citep{DaCunha2008, DaCunha2015} and enabling for a minimum stellar mass of 10$^6$ M${\odot}$ \citep[see Sec. 3.2 of][]{Maseda2017}. 
We used HST photometry from the UVUDF catalog of \cite{Rafelski2015} which comprises WFC3/UVIS {\itshape F225W}, {\itshape F275W, } and {\itshape F336W};  ACS/WFC  {\itshape F435W}, {\itshape F606W}, {\itshape F775W}, and {\itshape F850LP} and WFC/IR {\itshape  F105W},  {\itshape F125W}, {\itshape F140W,} and {\itshape F160W}.

The templates incorporated in \magphys\ do not include contributions from nebular emission. This does not strongly affect our SED fitting results as the strongest optical emission lines (\hbw, \oiiibpt\ and \haw) do not fall in any of the HST passband filters used in the fitting process for $z>2.9$. However, we note that the weaker \oiid\ can contaminate the {\itshape F160W} channel for $z\lesssim3.2$, and \lya\ itself falls in the {\itshape F606W} channel for the redshift range of interest in this work. Even if the \lya\ EWs are smaller that those of the Balmer lines, this could bias our stellar mass estimates to larger values in some cases \cite[cf.,][]{Stark2013}.
In addition, the parameters inferred from SED fitting can suffer a certain level of degeneracy as a uniform catalog with deep near-IR (e.g., {\it Spitzer}/IRAC) data points, redward of the HST {\itshape F160W} passband filter, is not yet publicly available. 
For the above reasons, we limited ourselves to using the physical properties, namely stellar mass, SFR, and sSFR, inferred from SED fitting to split the sample in two bins using the median value (dashed lines in all the panels).

Spectroscopic studies in a similar redshift range to the one considered in this work have been performed with VLT/VIMOS and Keck/DEIMOS \citep{Cassata2011, Dawson2007, Cassata2015, Pahl2020}. While the limiting magnitude at 1500 \AA, M$_{\rm UV}$  of these previous studies is roughly $-$18, the MUSE HUDF survey has enabled statistical studies of hundreds of fainter LAEs with UV magnitudes in the range of $-21.3 <$ M$_{\rm UV} < -15.8 $. Our observations probe fainter luminosities compared to other samples of LAEs at $z\simeq 2-3$ where the limiting UV magnitude is M$_{\rm UV} = -18$ \citep[e.g.,][]{Erb2014, Nakajima2018b} or M$_{\rm UV} =-19$ in the case of deep VANDELS \heii\ emitters \citep{Saxena2019}, or M$_{\rm UV} =-20$ for the sample of $\sim 1000$ Lyman-break galaxies from \cite{Shapley2003} \citep[see Fig. 1 of][]{Nakajima2018b}. 
Previous spectroscopic observations of targets with UV luminosities as faint as those observed in our sample have been mainly performed thanks to gravitational lensing \citep[e.g.,][]{Stark2014, Bina2016} and for small samples of tens of galaxies.

%-----------------------------
% Figure 1
%

%-----------------------------
% Figure 1b
  \begin{figure}[!h]
  \centering
   \includegraphics[width=8.8cm]{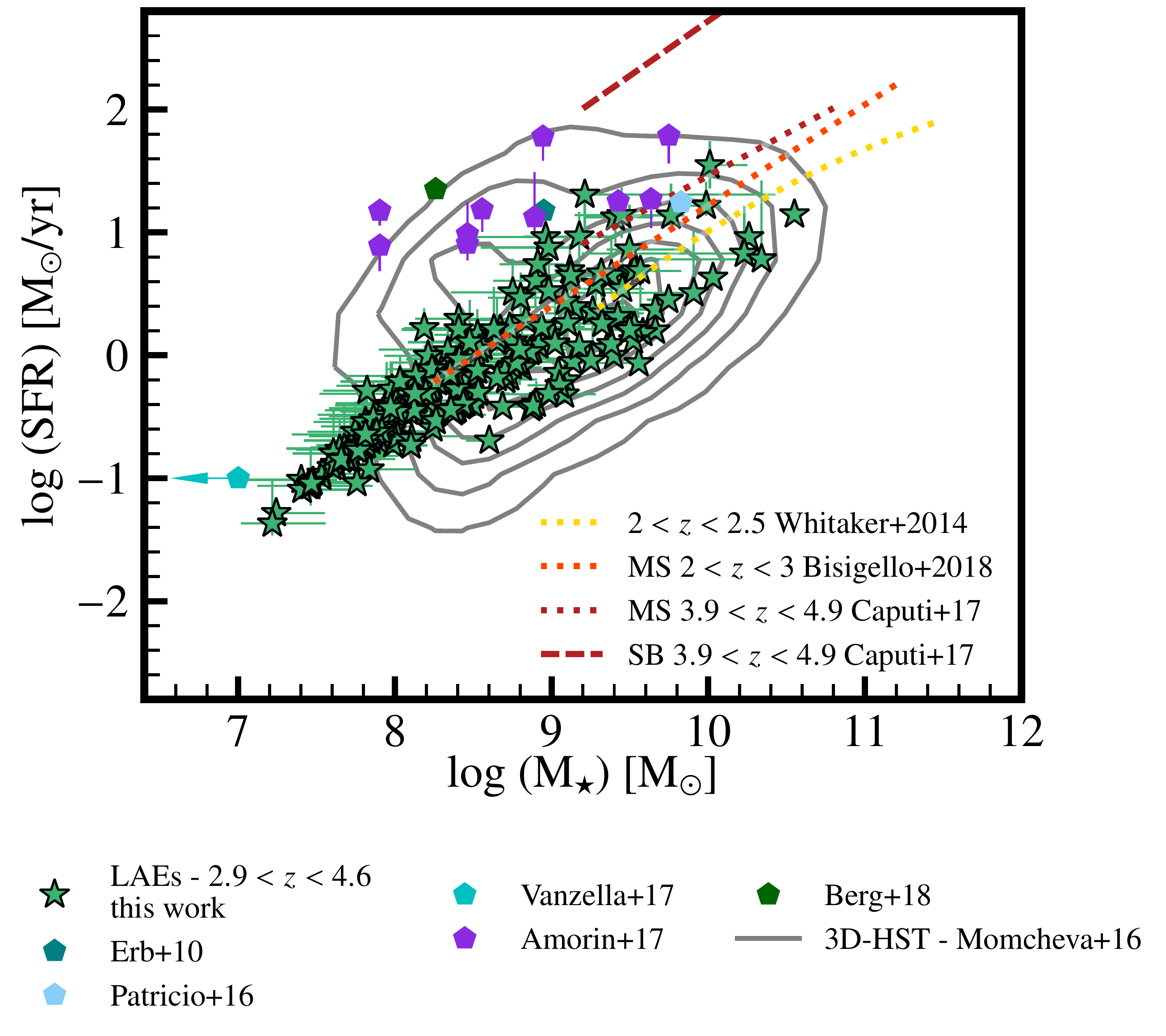}
   \caption{Distribution of the LAE sample (green stars) in the star formation rate-stellar mass (SFR-M$_{\star}$) plane. For comparison, values for $2.4<z<5.5$ galaxies in the 3D-HST survey public catalog by \cite{Momcheva2016} and data for $z \gtrsim2$ star-forming galaxies from \cite{Erb2010}, \cite{Patricio2016}, \cite{Amorin2017}, \cite{Vanzella2017}, and \cite{Berg2018} (see also Sect. \ref{sec:comparison}) are shown color-coded as labeled in the legend. The curves of the star formation sequence from \cite{Whitaker2014} at $2<z<2.5$, main sequence (MS) at $2<z<3$ from \cite{Bisigello2018}, and at $3.9<z<4.9$ from \cite{Caputi2017} and the starburst (SB) sequence at $3.9<z<4.9$ from \cite{Caputi2017} are shown for reference.}
            \label{Fig1b}%
   \end{figure}

Our sample does not include LAEs with such high luminosities ($L_{{\rm Ly}\alpha} \gtrsim 10^{43}$ erg s$^{-1}$) as, for example, wider field spectroscopic and narrow band surveys \citep[e.g.,][]{Gronwall2007, Ouchi2008, Cassata2011, Sobral2017, Matthee2017}, which all seem to host an AGN \citep[e.g.,][]{Konno2016, Sobral2018}. We measured \lya\ fluxes down to a few $\times 10^{-18}$ erg s$^{-1}$ cm$^{-2}$, similar to a few tens of sources from \cite{Rauch2008} and \cite{Cassata2011}, while many other surveys probed only the most luminous emitters with \lya\ fluxes larger than $10^{-17}$ erg s$^{-1}$ cm$^{-2}$ \citep[e.g.,][]{Ouchi2008, Sobral2017}. The FWHMs of \lya\ are comprised between $\approx 100$ km s$^{-1}$ and $\approx 640$ km s$^{-1}$, with a median value of $\approx 270$ km s $^{-1}$.
Our LAEs have a median value of the rest-frame \lya\ EW of $\approx 90$ \AA\ and include high \lya\ EW values, 34 (9) with EW$_{\lya} \geq 200 \, (400)$ \AA\ similar to the extreme emitters detected by \cite{Hashimoto2017a} at z$\approx 2$ \cite[see also, Section 8.1 of H17 and][]{Maseda2018}. 
The \lya\ EWs of our sample are $>10\, \AA$ and on average higher than those found in samples of high-z galaxies selected by photometric redshifts, that is,\ requiring a detectable continuum in several photometric bands \citep[e.g.,][]{Cassata2015, Cullen2020}. In consequence, many of our galaxies are much fainter in their continua than in other samples, which we exploit in this study. 

The UV continuum slope $\beta$ ($f_{\lambda} \propto \lambda^{\beta}$) measured in  the  rest-frame  wavelength interval of $\sim1700-2400\AA$ (Sect. 3.1 of H17) ranges from $\approx -3$ to 1, with a median value of $-$1.8. A significant fraction ($\approx 37\%$) of LAEs have blue UV slopes, $\beta \lesssim -2$, as those observed in dwarf galaxies at $z\approx2$ \citep[e.g.,][]{Stark2014} and at $z\simeq7$ \citep[e.g.,][]{Finkelstein2012, Bouwens2012, Bouwens2013}, suggestive of little reddening from dust. 
The low dust content of most of our LAEs is also supported by the low attenuation in the V-band, $A_{\rm V}$, inferred from the \magphys\ fitting tool: median (mean) values of 0.04 (0.11) mag.

This study probes some of the lowest stellar mass objects observed at the redshift range of interest, down to $\approx 1.6 \times 10^7 \,{\rm M}_{\sun}$ with SFRs in the range of $0.04< {\rm SFR} <35 \, {\rm M}_{\sun}\, {\rm yr}^{-1}$. Figure \ref{Fig1b} shows the distribution of our MUSE HUDF LAEs in the star formation rate-stellar mass (SFR-M$_{\star}$) plane, which overall do not populate areas strongly above the so-called main sequence of galaxies \citep{Noeske2007, Whitaker2012} as other $z>2$ line emitter galaxies presented in the literature \cite[e.g.][]{Erb2010, Amorin2017, Berg2018}. For comparison, we show the distributions of the stellar masses and SFR of $2.4 < z < 5.5$ galaxies from the public catalog of the 3D-HST survey \citep{Momcheva2016}, which also include the targets of spectroscopic surveys like VANDELS \citep{McLure2018}. The sSFR, $-9.6 < {\rm log_{10}(sSFR /yr^{-1})} < -7.9$, are on average less extreme than those (${\rm log_{10}(sSFR /yr^{-1})} \geq -7.7$) found in local and high-$z,$ metal-poor star-forming galaxies and Lyman continuum (LyC) leaker candidates \citep[see column 6 of Table 1 from][]{Plat2019}.

\subsection{Spectral extractions}\label{sec:extractions}

Three different methods for spatially integrating the data cube and creating one-dimensional spectral extractions are presented in I17 (Sect. 3.1.3): namely the unweighted sum, the white-light weighted, and the PSF-weighted. In this work, we stack PSF-weighted spectral extractions of the 220 LAEs described in Sect. \ref{sec:sample}, as they offer the advantage of a higher S/N of the extracted one-dimensional spectra, a reduced contamination from neighboring objects, and flux conservation that enables easy data comparison. Moreover, the PSF-weighted version is the reference extraction (\texttt{REF$\_$SPEC} in I17) used to compute the spectroscopic redshift for the majority of the sample (215/220). The assignment of a weighted optimal spectra as reference extraction in the MUSE HUDF DR1 catalog depends, as described in Sect. 3.1.3 of I17, on the galaxy size as computed by \texttt{SExtractor} \citep{Bertin1996} in the HST {\itshape F775W} images from the UVUDF catalog \citep{Rafelski2015}. The PSF-weighted spectrum is adopted for objects with FWHM $<0.7$".

In Sect. \ref{sec:subsamples}, we group our sample in two bins for each of the properties described in Sect. \ref{sec:properties}, including \lya\ properties. As noted in Sect. 3.3 of I17, possible biases on flux measurements introduced by the choice of a weighted extraction could be important, particularly to avoid losing spatially extended line emission, as in the case of the resonant \lya\ line. To obtain accurate estimates of the total \lya\ emission, we used the \lya\ fluxes and EWs recomputed by applying the CoG method \citep[e.g., Sect. 5.3 of][]{Leclercq2017} accounting for the likely extended emission whose origin is currently a subject of intense discussion \citep[e.g.,][]{Wisotzki2016, Leclercq2017, Kusakabe2019}. Given that the other UV lines have higher ionization potential, their emission is expected to be less extended than \lya. We visually inspected the \lya\ narrow band images to check against radiation from close companions affecting our \lya\ CoG fluxes. 
We checked that by grouping our sample considering the \lya\ properties (flux, luminosity, EWs, and FWHM) computed on the weighted optimal extractions, rather than using the CoG method, the trends described in Sect. \ref{sec:results} and our conclusions (Sect. \ref{sec:conclusions}) remain unchanged.

%-----------------------------
% Figure 2
  \begin{figure*}
  \centering
   \includegraphics[width=14.5cm]{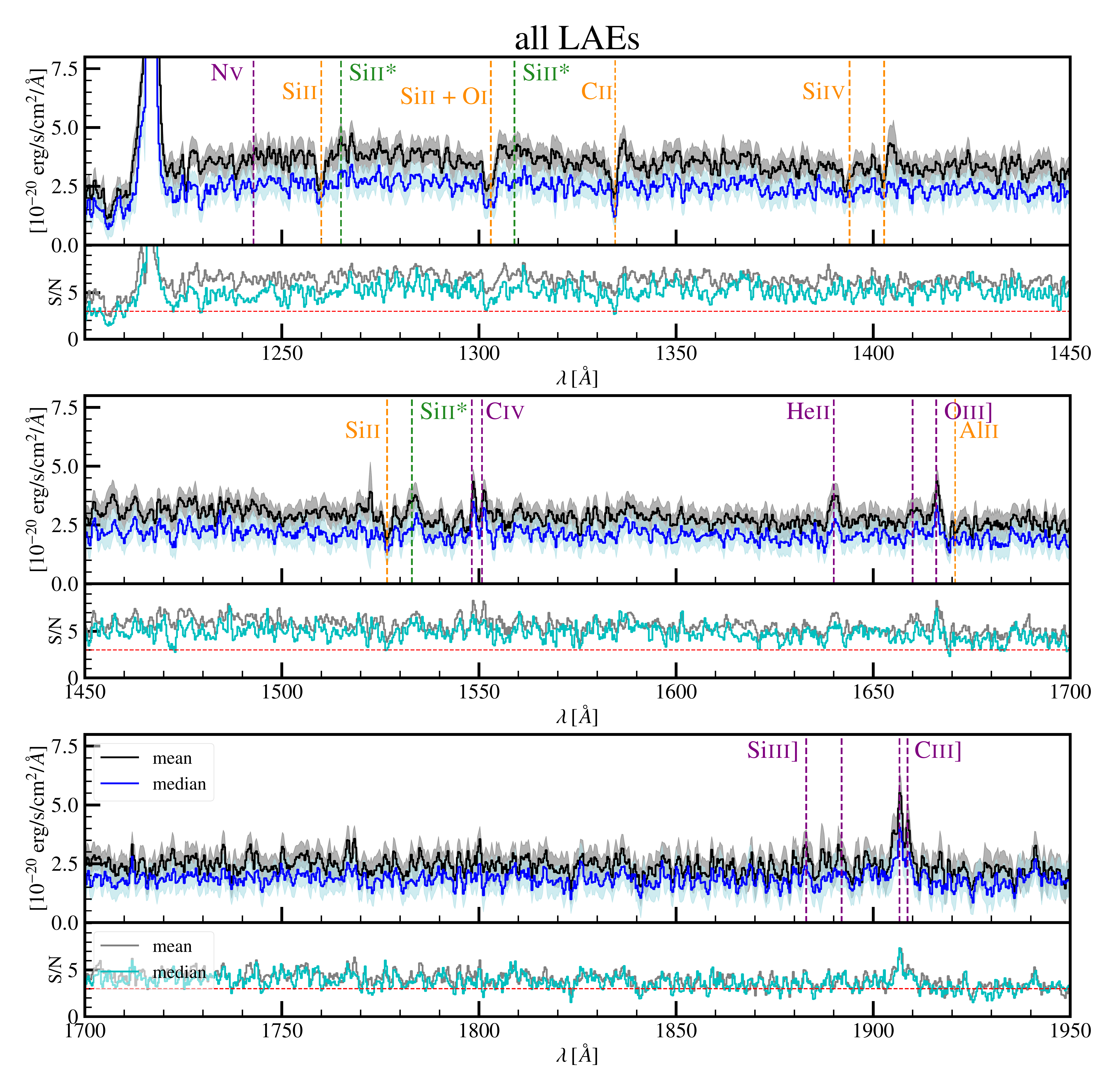}
   \caption{Mean and median spectra of all 220 LAEs described in Sect. \ref{sec:sample} shown in black and blue with the error displayed via gray and light-blue shaded areas, respectively. The vertical lines indicate the rest-frame wavelengths of some of the main spectral features: nebular emission lines (purple), ISM absorption (orange), and fine-structure transitions (green). The bottom panels show the S/N per pixel (gray and cyan curves for mean and median spectra, respectively) with the dotted red horizontal line indicating S/N =3.}
           \label{Fig2}%
   \end{figure*}

\section{Spectral stacking}\label{sec:stacking}

\subsection{Stacking methodology}\label{sec:stack_methodology}

We statistically combined multiple galaxy spectra to obtain the composite spectrum of different LAE subsamples (defined in Sect.~\ref{sec:subsamples}) and adopted a bootstrapping method to compute the error of the stacked spectra. For most of our sources, the redshift was measured from the \lya\ line (see I17), which is often shifted by few hundred km s$^{-1}$ compared to the systemic redshift \citep[e.g.,][]{Hashimoto2013, Song2014, Erb2014, Trainor2015, Henry2015, Hashimoto2015}. We therefore computed new redshfits using the empirical correlation between the velocity offset of the \lya\ peak and the \lya\ FWHM proposed by \citet[][hereafter V18, Eq. 2]{Verhamme2018}, assuming that the systematic errors are negligible compared to the statistical errors. The uncertainties related to these prescriptions are discussed in Sect. ~\ref{sec:lya_correction}.
We then shifted the spectra of the LAEs to their corrected redshifts and re-binned them to a linear sampling of 0.3~\AA\ (corresponding to  the 1.25~\AA\ sampling of MUSE at $z\approx3$).
The spectra were averaged over 150 bootstrap iterations in order to compute the noise associated with the stacked spectrum.

%-----------------------------
% Figure 3
  \begin{figure*}
  \centering
   \includegraphics[width=16.cm]{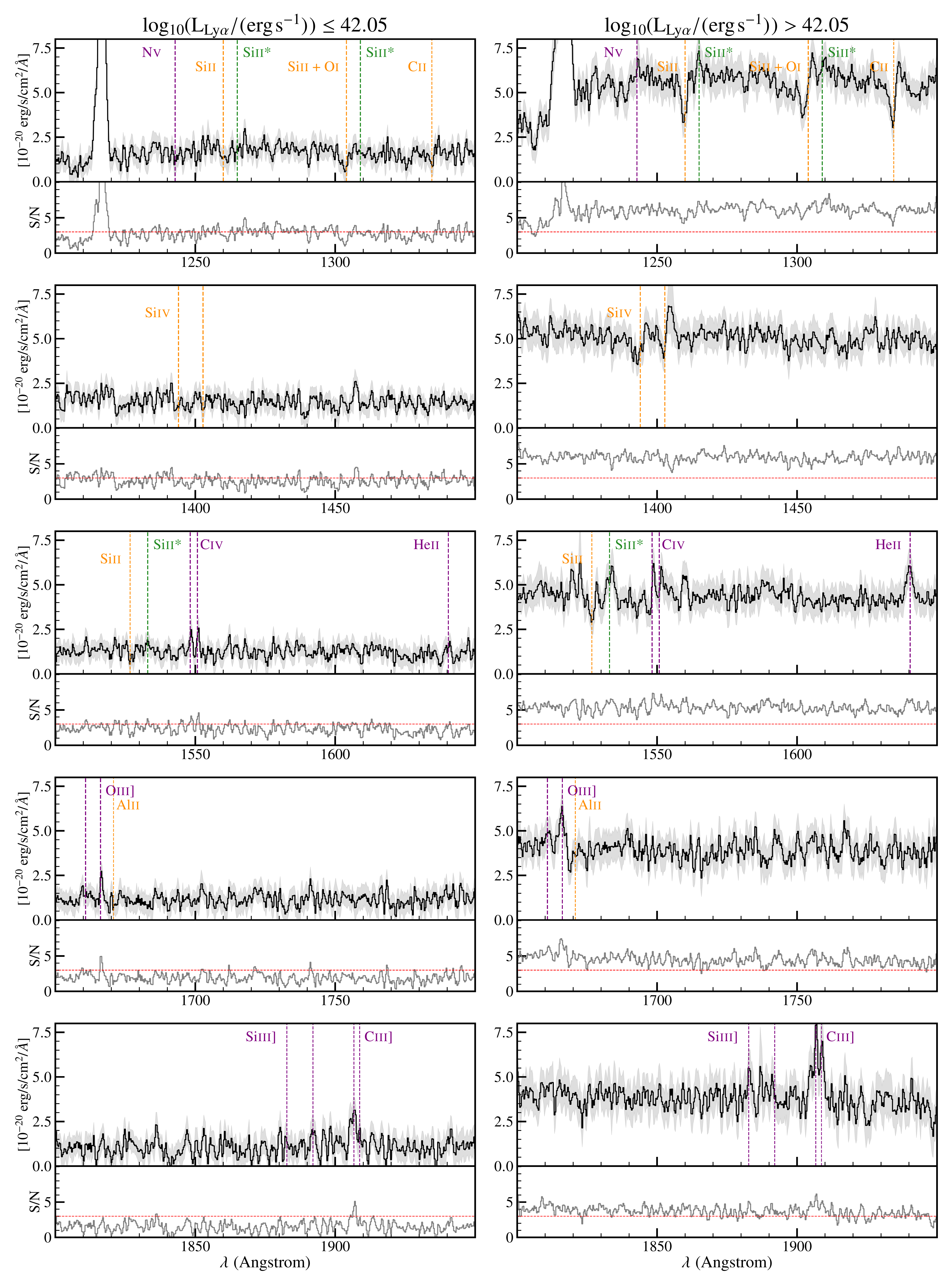}
   \caption{Mean spectra of LAEs with log$_{10}$(L$_{{\rm Ly\alpha}}$ / (erg s$^{-1}$)) $\leq 42.0$ and $> 42.05$ (left and right, respectively) are shown in black in the top panels, with the error budget displayed via the gray shaded regions. The median values of log$_{10}$(L$_{{\rm Ly\alpha}}$  / (erg s$^{-1}$)) for the two subsamples are 41.86 and 42.35, respectively. 
   The vertical lines indicate the rest-frame wavelengths of some of the main spectral features: nebular emission lines (purple), ISM absorption (orange), and fine-structure transitions (green). The bottom panels show the S/N per pixel (gray curves) with the dotted red horizontal line indicating S/N =3.}
            \label{Fig3}%
   \end{figure*}

There are different possible approaches to combine the spectra from a stack. We used simple unweighted averaging to obtain spectra that are as representative of our LAE subsamples as possible. We decided against the application of any flux-related weighting schemes (e.g., by the inverse variances) because such schemes invariably lead to composites dominated by the few brightest spectra.
We compare mean and median stacks, as the brighter and more massive LAEs or galaxies with stronger lines could dominate the signal in the mean spectra of some LAE subsamples. We did consider median stacking in addition to just the mean, which reduces the influence of the brightest and most massive galaxies even further, but at the expense of obtaining a lower SNR in the median composites.  Overall, we found that the EWs measured in mean and median stacks are consistent within one standard deviation. The line ratios differ by up to 0.1--0.2 dex comparable with the error measurements of the ratios.  The mean and the median stacked spectra of the full sample of LAEs are shown in Fig. \ref{Fig2} (black and blue curves, respectively).

We note that the MUSE spectra have not been corrected for dust attenuation due to the complexity of properly modeling the dust embedded in different emitting sources (stars and gas) and to the poor constraints available at high $z$ \citep[e.g.,][and references therein]{Chevallard2013, Schaerer2015, Reddy2016, Buat2018}. Most of the LAEs in our sample have blue UV slopes, suggesting a low dust content. Moreover, SED fitting to the broad-band photometry indicates a median attenuation of 0.11 mag in the $V$-band. If we take this as proxy for the continuum and nebular attenuation, we can assume a minimum impact from dust on the observed spectra of our LAEs.

\subsection{Stacked spectra of LAEs}\label{sec:subsamples}

For each of the properties described in Sect.~\ref{sec:properties}, we divided our sample of LAEs in two subsamples adopting as thresholds the median values of the distributions (black vertical lines in the histograms of Fig.~\ref{Fig1}). This ensures roughly the same number of LAEs for each subsample. It is worth noting that none of the spectra of the individual 220 LAEs have S/N$>3$ detections of \ciii\ or any other UV emission lines apart from \lya\ (Sect. \ref{sec:results}). 

We then computed the stacked spectra of all the subsamples in order to study and compare their average spectral properties. 
As an example, Fig.~\ref{Fig3} shows the stacked spectra of fainter and brighter LAEs with log$_{10}$(L$_{{\rm Ly\alpha}}$ / (erg s$^{-1}$)) $\leq 42.0$ and $> 42.05$ (left and right, respectively). All the average spectra of the other subsamples are shown in Appendix \ref{app:A}, and their emission line properties are discussed in Sect.~\ref{sec:results}.

We note that the limited number of sources in our sample prohibits additional binning, and namely splitting them into three or more subsamples. In particular, for lower luminosity and lower mass LAEs, we did not detect emission lines with S/N $> 2.5,$ even though smooth transitions between the stacked spectra could be visually apparent.

\subsection {Velocity offset correction}\label{sec:lya_correction}

The redshift of the MUSE HUDF galaxies at $z>3$ is often determined solely via the \lya\ line, which, due to its resonant nature, is often offset by a few hundred km s$^{-1}$ compared to the systemic redshift \citep[e.g.,][]{Hashimoto2015}. \cite{Verhamme2018} propose two prescriptions to recover the systemic redshift from the observed correlations between the separation of the peaks or the FWHM of the \lya\ line and the velocity offset of the red peak ($V^{{\rm red}}_{{\rm peak}}$) with respect to the systemic redshift. 
As not all the LAEs of our sample show a double-peaked \lya\ line profile, we adopted the latter correlation (Eq. 2 of V18). 
Other empirical prescriptions to recover the systemic redshift of LAEs have been proposed in the literature, such as the relationships between the velocity offset and the \lya\ EW proposed by  \cite{Adelberger2003} and \cite{Nakajima2018b}.
The corrected redshifts computed by adopting these two prescriptions differ at maximum by 10\% of the uncertainty associated with the redshift computed with Eq. 2 of V18. 

\begin{figure}[!h]
  \centering
   \includegraphics[width=8cm]{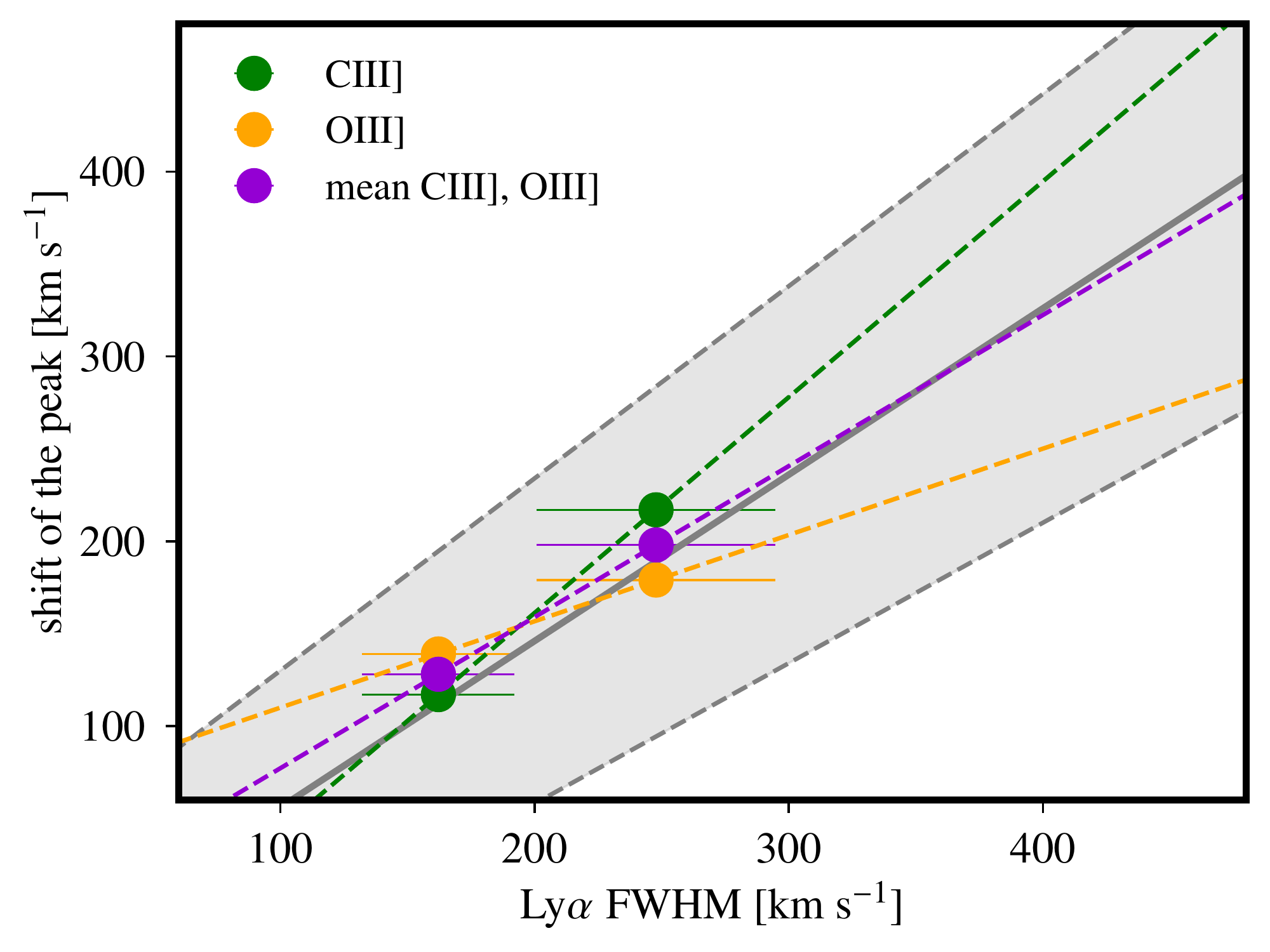}
   \caption{Correlations between shift of \lya\ peak and \lya\ FWHM obtained from \lya\ FWHM subsamples of our LAEs. These correlations were obtained considering the rest wavelengths of \ciii\ (green line) and \oiiisf\ (orange line) as reference wavelengths and their mean shift (purple line). In gray, Eq. 2 of V18, obtained from object-by-object measurements, and the associated scatter (gray shaded area).}
            \label{Fig4}%
   \end{figure}

By inspecting our stacked spectra, we find some emission line peaks to be redshifted compared to the rest wavelength of the line (Figs. \ref{Fig3} and in Appendix \ref{app:A}). For a given stacked spectra, this shift is not the same for all the emission lines, but it varies from 10 to 50 km s$^{-1}$, reaching values up to 100 km s$^{-1}$ only when the line has S/N$<2.5$. This may suggest that a unique relation is not optimal for all the sources. We consider the subsample split in \lya\ FWHM (Fig. \ref{FigA3}) and measure the velocity shift of the peak using the rest wavelengths of the \ciii\ and \oiiisf\ collisionally excited lines as reference wavelengths. In this way, we obtain, for each line, two points to anchor the relation between the \lya\ FWHM and shift of the peak for our LAEs (Fig. \ref{Fig4}, green and orange lines). We also consider the mean shift with respect to the two ISM emission features (purple line), which is fully consistent with the V18 relation (in gray). Moreover, the relation obtained using \oiiisf\ as reference line has a slope very similar to the relation found by \cite{Muzahid2020} obtained from stacking circumgalactic medium (CGM) absorption profiles. It is remarkable how different methods, either object-by-object-based, or averaged over a larger number of sources, provide similar correlations, all within the scatter of the V18 relation. 

Given the uncertainty in the systemic redshift of our LAEs, the EWs of the stack are always underestimated compared to true values that one would measure if the correct systemic redshift were known. To investigate to what extent the uncertainty related to the V18 correction affects the spectral measurements of the emission lines presented in Sect.~\ref{sec:nebular_lines}, we computed two simulated, idealized spectra with constant continuum and FWHMs of 150 and 300 km s$^{-1}$, respectively.
We created 150 copies of these spectra using the same sampling of 0.3\AA\ as for the LAEs and added random Gaussian noise to each pixel, emulating the continuum S/N of our stack (S/N = 5). We then randomly shifted the line center according to Eq. 2 of V18 (assuming a value for \lya\ FWHM of 300 km s$^{-1}$, close to the median value of our sample) and stacked these idealized spectra with the same method adopted for our real sample of LAEs (Sect. \ref{sec:stack_methodology}). We find that the EWs of the these stacks can be underestimated between 15 and 20\% compared to the values of the idealized case. 
This means that the line EWs measured on the spectra of our LAEs and reported in Table \ref{tab:EW} can be up to $\approx$20\% higher than the tabulated value.

%-----------------------------
% Figure 4

\section{Results}\label{sec:results}
      
The mean and median stacked spectra of all the LAEs (Fig. \ref{Fig2}) show features in emission and absorption. These include nebular emission lines, like \oiiisf\ and \ciii, and absorption features, such as  \oisiliit\ and \siliib\ (hereafter \oisilii\ and \silii, respectively). The latter have been identified as uncontaminated tracers of interstellar absorption in the UV spectra of star-forming galaxies \citep{Vidal-Garcia2017}.
 The other interstellar absorption features that are clearly visible in Fig. \ref{Fig2}, \siliia\  and \ciia\ (hereafter \cii), can be contaminated by nebular emission \citep{Vidal-Garcia2017}. The \silivl\ (hereafter \siliv) ISM absorption feature, which, along with \civ, traces the highly ionized ISM, is also contaminated by stellar wind features. The \heii\ and \civ\ emission lines are blends of stellar photospheric absorption/winds and nebular emission. In addition, \civ\ is a resonant line affected by absorption from gas in the ISM and CGM surrounding galaxies. Most of these features have been observed in the stacked spectra of $z>2$ galaxies. Interestingly, while the the stacked spectra of \cite{Shapley2003} and \cite{Saxena2019} show \civ\ absorption or P-Cygni, the mean \civ\ doublet of our LAEs (Fig. \ref{Fig2}) is in emission.

This work focuses on measuring the UV nebular emission lines in the mean spectra of the different LAE subsamples (Sect.~\ref{sec:nebular_lines}) and studying how these features vary with the observational and physical properties of these LAEs (Sect.~\ref{sec:discussion}). The stacked spectra of the LAE subsamples show a large variety of nebular emission lines, absorption features, and fine structure transitions, as shown in Figs. \ref{Fig2}, \ref{Fig3}, and in Appendix \ref{app:A}. None of the 220 LAEs have UV emission lines other than \lya detected in their individual spectra with S/N > 3. The only exception is ID3621, in the \mosaic\ field, for which the \heii\ line is detected with S/N$>2.5$ \citep{Nanayakkara2019}. 

In the stacks of Figs. \ref{Fig3} and Appendix \ref{app:A}, we identify some of the main UV nebular emission lines that are currently detected in the spectra of high-$z$ or local, metal-poor star-forming galaxies \citep[e.g.,][]{Erb2010, Stark2014, Berg2016,Berg2018, Berg2019, Patricio2016, Vanzella2016,Vanzella2017, Nakajima2018b}. Namely, we detected the collisionally excited \oiiisf\ and \ciii\ doublets, and, in few cases, \siliii, as well as the \heii\ emission feature. In addition, we observed different profiles of the \civ\ resonant doublet, as discussed in Sect.~\ref{sec:civ}. Depending on the observed and physical properties of the LAEs, the stacked spectra of several subsamples exhibit absorption features and fine-structure transitions (see Sect.~\ref{sec:discussion}). A quantitative study of these features would require a complex modeling that combines stellar continuum, nebular emission and resonant scattering through the ISM and CGM. This is beyond the scope of the current analysis and will be the subject of future works.

\subsection{\textnormal{\oiiisf, \ciii, \siliii,} and \heii\ lines}\label{sec:nebular_lines}

We computed EWs of \heii, \oiiisf, \ciii,\ and \siliii\ by performing a Gaussian fit to the portion of the spectrum that contains the emission line, fixed to the systemic redshift; in particular, we fit the line doublets as a sum of two Gaussian functions. The continuum is defined by calculating the mean flux within a window of 100 \AA\ around the emission lines, excluding the central 10 \AA\ around the rest wavelengths of the line features. 
We note that we did not subtract the stellar continuum during this fitting procedure. This should not strongly affect the collisional lines such as \ciii\ and \oiiisf,\ but multiple mechanisms are known to contribute to the recombination \heii\ line, which has both a nebular and stellar origin. For this reason, the width of the \heii\ line is free to vary in the fit. We found its FWHM to be from 0.6 to 2.5 times that of \ciii\ and \oiiisf, consistent with the results from \citet[][Fig. 16]{Nanayakkara2019}. In our case, the possibility of constraining the potential contribution from Wolf-Rayet stellar winds to the total \heii\ flux is hampered by the low S/N. Given the current limitations of theoretical models in correctly reproducing the nebular \heii\ and the uncertainties in modeling the continuum emission from massive stars \citep[e.g.,][]{Senchyna2017, Berg2018, Nanayakkara2019, Plat2019}, here we simply provide the integrated fluxes obtained from the Gaussian fitting, without including additional uncertainties in the decomposition of the different mechanisms that can power the \heii\ feature.  

\begin{figure*}[!h]
  \centering
   \includegraphics[width=20cm]{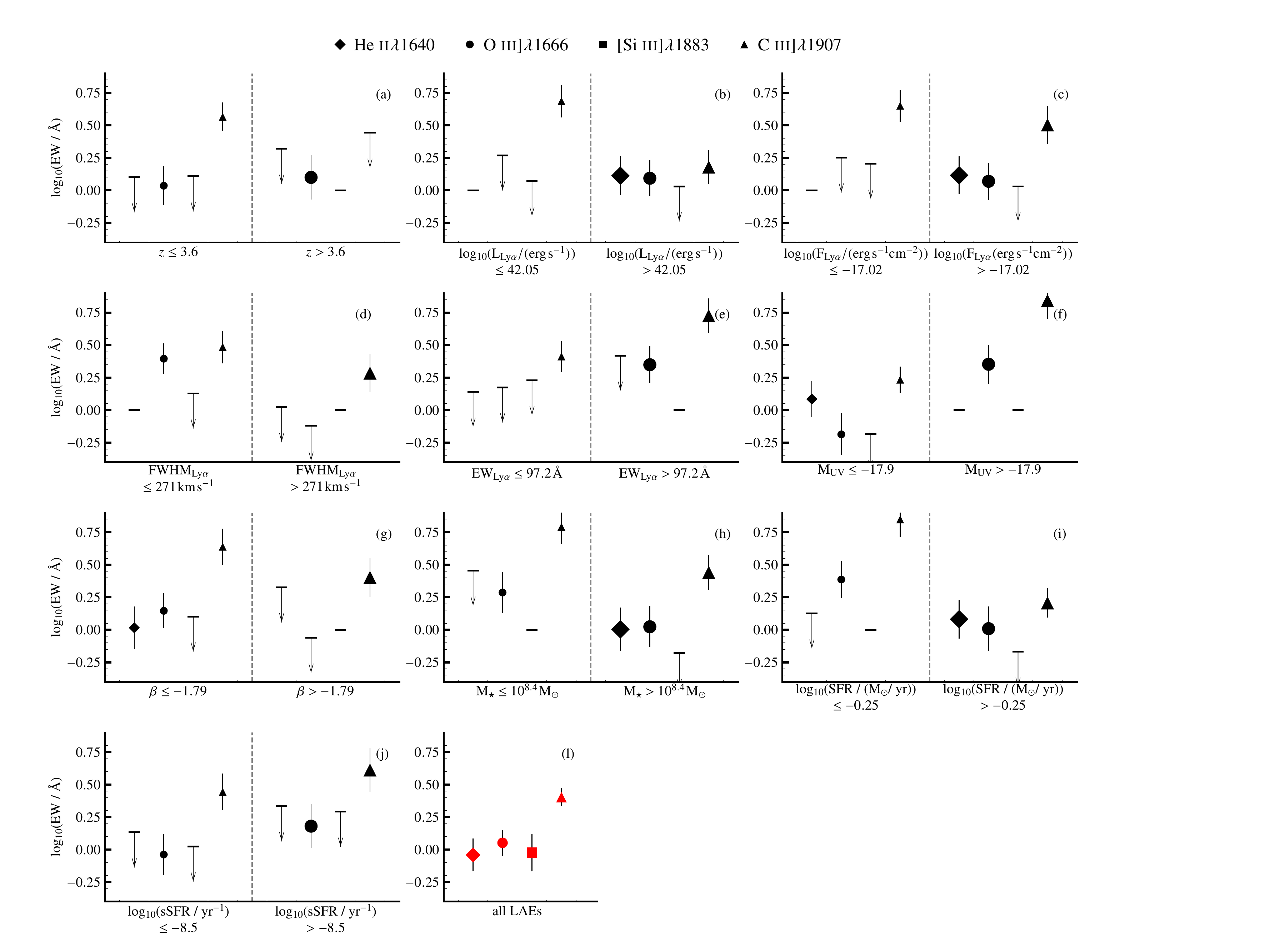}
   \caption{EWs of \heii, \oiiir, \siliiib,\ and \ciiib\ for the different subsamples of MUSE HUDF LAEs (x-axis). The symbols refer to different emission lines, as labeled in the legend. Smaller and larger symbol sizes indicate the subsample of LAEs whose properties are lower and higher than the median value, respectively. Empty symbols with downward arrows indicate upper limits, while non-detections are shown by the black dash. Red symbols show the EWs for all 220 LAEs in the sample.}
            \label{Fig5}%
   \end{figure*}

The values of the EW of lines detected with S/N $>2.5$ in the composite spectra  of the different subsamples (and of the complete 220 LAEs sample) are reported in Table \ref{tab:EW}. In some cases, and in particular for the faintest subsamples, even if the emission features are clearly visible in the stacked spectra, the noise prevents S/N $>2.5$ line detections. The EW values of the lines with S/N $<2.5$ are reported as 1-$\sigma$ upper limits in Table \ref{tab:EW}. No value is reported in the event of non-detection, meaning when the line is embedded within a highly noisy continuum level.
The nebular-line EWs for the different LAE properties are shown in Fig. \ref{Fig5}, while a comparison with other measurements from the literature and line ratios are shown in Fig. \ref{Fig7} and Fig. \ref{Fig8} (Sect.~\ref{sec:comparison}). 

The \ciiib\ blue component of the \ciii\ doublet is detected in almost all the stacks except for $z>3.6$. Out of the five stacked spectra where both components are detected, four show a \ciiib\ / \ciiir\ ratio in excess of the value expected in the low-density limit (1.53), implying electron densities $\gtrsim10^3 {\rm cm^{-3}}$ \citep[see section 4.1 of][]{Maseda2017}. We detected the \oiiir\ component of the \oiiisf\ intercombination doublet in all the stacks except for the log$_{10}$(L$_{{\rm Ly}\alpha} / ({\rm erg \, s^{-1}})) \leq 42.05$, log$_{10}$(F$_{{\rm Ly}\alpha} /({\rm erg s^{-1} cm^{-2} })) \leq -17.02$, \lya\ FWHM $>$ 271 km s$^{-1}$, \lya\ EW $\leq 97.2$ \AA\ and $\beta >-1.79$, while the typically weaker blue component of the doublet is never detected with S/N $> 2.5$. The \heii\ line is observed in the stacks of LAEs with more intense \lya\ emission, M$_{\rm UV} \leq -17.9$, bluer UV slope, and a higher stellar mass and SFR. The \siliii\ doublet is detected with S/N$>2.5$ only in the stack of the total sample, but not in those of the subsample split. In column 6 of Table \ref{tab:EW}, we report 1-$\sigma$ upper limits for the \siliiib\ component, as this is stronger than \siliiir\ for electron densities $<10^5\, {\rm cm^{-3}}$.

%-----------------------------
% Figure 5

\begin{table*}
   \centering
    \caption{Details of the MUSE HUDF LAE subsamples and emission-line EWs measured from the stacked spectra}
    \label{tab:EW}
    \begin{tabular}{lccccccc} % <-- Alignments: 1st column left, 2nd middle and 3rd right, with vertical lines in between
      No. & Subsamples & \# of LAE & \heiil\ &  \oiiir\ & \siliiib\ & \ciiib\ & \ciiir\ \\
      \hline 
      \hline 
      00 & all LAEs & 220 & $0.91\pm0.26$ & $1.13\pm0.26$ & $0.95\pm0.32$ & $2.53\pm0.41$ & $1.43\pm0.39$ \\
      \hline 
      01 & $z\leq3.6$ & 110 & $<1.27$ & $1.08\pm0.37$ & $<1.28$ & $3.68\pm0.93$ & $<1.53$ \\ 
      02 & $z>3.6$ & 110 & $<2.09$ & $1.26\pm0.50$ & -- & $<2.78$ & $<1.50$   \\ 
      \hline
      03 & log$_{10}$(L$_{{\rm Ly\alpha }}/({\rm erg\, s^{-1}})) \leq 42.05$ & 110 & -- & $<1.86$ & $<1.18$ & $4.85\pm1.40$ & $<1.67$ \\ 
      04 & log$_{10}$(L$_{{\rm Ly\alpha }}/({\rm erg\, s^{-1}})) > 42.05$ & 110 & $1.30\pm0.45$ & $1.24\pm0.39$ & $<1.07$ & $1.51 \pm0.46$ & $1.52\pm0.55$   \\ 
      \hline
      05 & log$_{10}$(F$_{{\rm Ly\alpha }} / ({\rm erg\, s^{-1} cm^{-2}})) \leq -17.02$ & 110 & -- & $<1.79$ & $<1.60$ & $4.48\pm1.25$ & $<1.18$ \\ 
      06 & log$_{10}$(F$_{{\rm Ly\alpha }} / ({\rm erg\, s^{-1} cm^{-2}})) > -17.02$  & 110 & $1.31\pm     0.44$ & $1.17\pm0.38$ & $<1.20$ & $3.19\pm1.07$ & $<1.66$   \\ 
      \hline
      07 & FWHM$_{{\rm Ly\alpha }}$ $\leq$ 271 km s$^{-1}$ & 99 & -- & $2.49\pm0.69$ & -- & $3.18\pm0.87$ &$ <3.53$\\
      08 & FWHM$_{{\rm Ly\alpha }}$ $>$ 271 km s$^{-1}$ & 99 & $<1.06$  & $ < 0.76$   & $ <1.44  $  & $1.92\pm0.65$ & $<1.46$\\
      \hline
      09 & EW$_{{\rm Ly\alpha}} [\AA] \leq 97.2$ & 102 & $<1.38$ & $<1.50$ & $<1.05$ & $2.58\pm0.72$ & $<1.04$   \\ 
      10 & EW$_{{\rm Ly\alpha}} [\AA] > 97.2$ & 102 & <2.63 & $2.23\pm0.73$ & $<1.70$ & $5.32\pm1.64$ & $<3.23$  \\ 
      \hline
      11 & M$_{\rm UV} \leq-17.9$ & 102 & $1.22\pm0.39$ & $0.65\pm0.24$ & $<0.65$ & $1.71\pm0.40$ & $1.01\pm0.39$  \\ 
      12 & M$_{\rm UV} >-17.9$ & 102 & -- & $2.25\pm0.78$ & -- & $6.96\pm2.28$ & $<3.29$  \\ 
      \hline
      13 & $\beta \leq-1.79$ & 102 & $1.03\pm0.39$ & $1.40\pm0.43$ & $<1.27$ & $4.36\pm1.39$ & $2.44\pm0.98$   \\ 
      14 & $\beta >-1.79$ & 102 & $<2.12$ & $<0.87$ &  -- & $2.53\pm0.87$ & $<0.81$ \\ 
      \hline
      15 & M$_{\star}$ [M$_{\odot}$] $\leq 10^{8.4}$ & 111 &$ < 2.85$ & $1.94\pm0.71$ & -- & $6.20\pm1.83$ & $ <3.40 $   \\ 
      16 & M$_{\star}$ [M$_{\odot}$] $> 10^{8.4}$ & 109 & $1.01\pm0.39$ & $1.06\pm0.38$ & $<0.66$ & $2.76\pm0.85$  & $<1.17$  \\ 
      \hline
      17 & log$_{10}$(SFR / (M$_{\odot}$/ yr)) $\leq -0.25$ & 110 & $< 0.90$ & $2.51\pm0.82$ & -- & $6.64\pm2.04$ & $<2.47$  \\ 
      18 & log$_{10}$(SFR / (M$_{\odot}$/ yr)) $ > -0.25$ & 110 & $1.21\pm0.42$ & $1.03\pm0.38$ & $<0.7$ & $1.61\pm0.42$ & $<0.76$ \\       
      \hline
      19 & log$_{10}$(sSFR / yr$^{-1}$) $\leq -8.5$ & 94 & $<0.92$ & $0.89 \pm 0.34 $ & $<0.65$ & $2.70\pm 0.85$ & $1.20\pm0.52$  \\
       20 & log$_{10}$(sSFR / yr$^{-1}$) $ > -8.5$ & 126 & $<1.40$ & $1.56\pm0.61$ & $<0.61$ & $3.84 \pm 1.44$ & -- \\ 
      \hline
      \hline
    \end{tabular}
    \tablefoot{Columns, from left to right: reference number of the stacked spectra, binning criterion of the LAE subsample, number of LAEs in each subsample, EW values or 1-$\sigma$ upper limits of \heii, \oiiir, \siliiib, \ciiib,\ and \ciiir\ lines. All EWs are rest frames and in \AA.}
\end{table*}

\subsection{CIV doublet and absorption features}\label{sec:civ}

\begin{figure*}[!h]
  \centering
   \includegraphics[width=17cm]{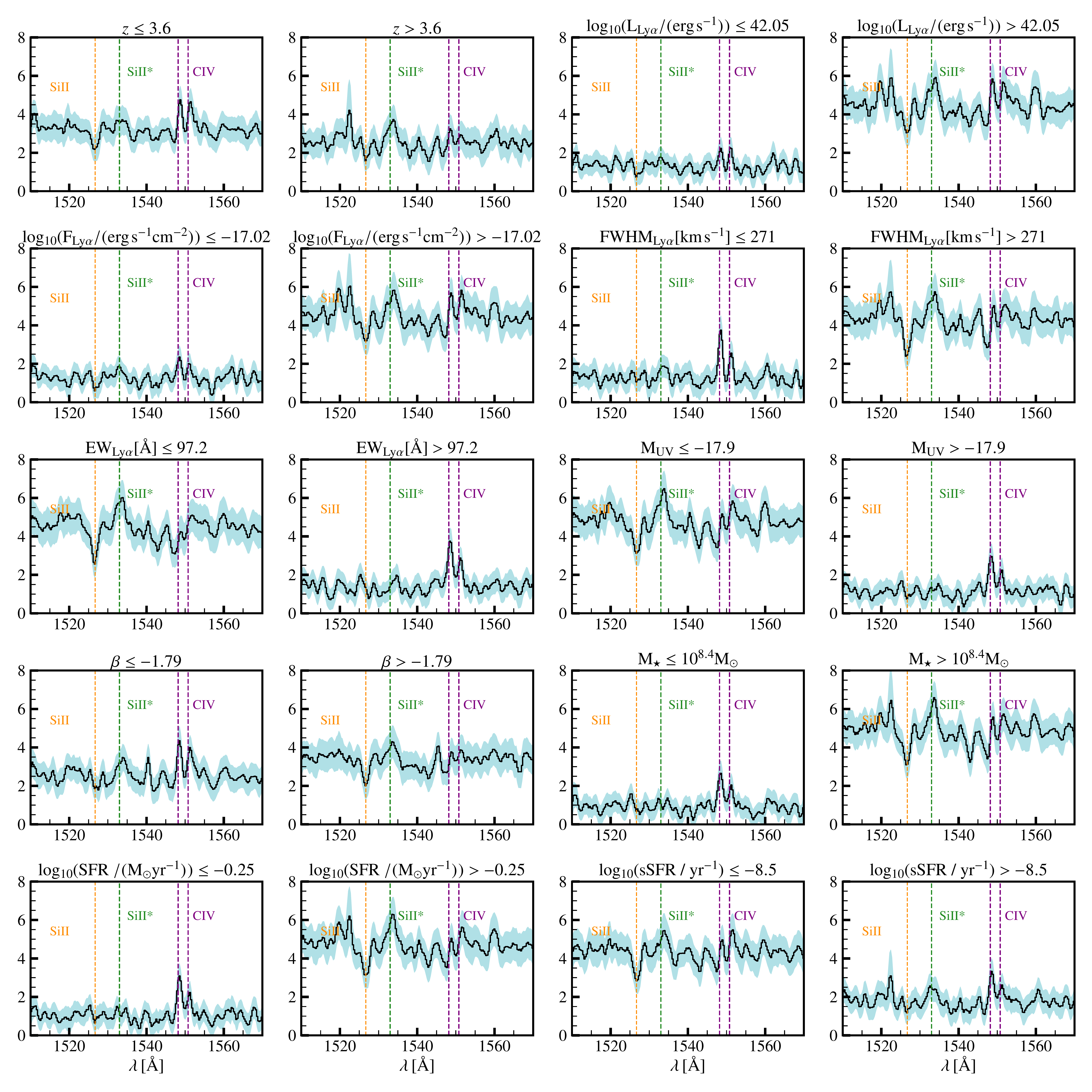}
   \caption{Zoom-in around the rest wavelengths of the \civd\ doublet (purple vertical lines) for the different averaged spectra (black lines) of the MUSE HUDF LAEs (as described in the title of each panel). The light blue shading indicates the noise. The other vertical lines indicate the rest-frame wavelengths of the \siliib\ ISM absorption feature (orange) and the \siliic\ fine-structure transition (green).}
            \label{Fig6}%
   \end{figure*}

Figure \ref{Fig6} shows the \civ\ resonant doublet, the \silii\ interstellar absorption, and the \siliic\ (hereafter \siliis) fine-structure transition for the composite spectra of the LAE subsamples. 
Multiple physical processes can contribute to the observed \civ\ spectral profile, like stellar winds in massive O and B stars, nebular emission and absorption in the ISM and CGM \citep[e.g.,][]{Vidal-Garcia2017, Byler2018, Berg2018}. Producing \civ\ emission requires photons with energies ($> 47.9$ eV) associated with hard ionizing radiation fields from massive stars, AGN, and radiative shocks. The \civ\ doublet is currently receiving a great deal of attention, as it is one of the strongest UV lines measured at high $z$ \citep[e.g.,][]{Stark2015b, Mainali2017, Schmidt2017, Vanzella2017, Berg2018} and in local metal-poor galaxies \citep[e.g.,][]{Senchyna2017, Senchyna2019, Berg2019}. Disentangling the multiple physical mechanisms contributing to the \civ\ profile requires a proper modeling of the stellar continuum, as well as of the resonant scattering through the medium within and around galaxies. A self-consistent modeling of photon scattering is important for the interpretation of the shapes of the interstellar absorption features, for example, \silii, and will be the subject of another work (Mauerhofer et al., in prep). In this work, we limit ourselves to a qualitative discussion of the appearance of the \civ\ feature and of the main absorption features.

We observed \civ\ with a P-Cygni profile in the stacked spectra of LAEs with higher \lya\ luminosity and flux, lower \lya\ EW, brighter UV magnitudes, redder UV slopes, higher masses, and higher SFRs (Fig. \ref{Fig6}). The origin of this P-Cygni profile can be associated with the presence of stellar winds from OB stars and depends both on the stellar metallicity and the relative fractions of O and B stars. Instead, the average spectra of LAEs with lower \lya\ FWHM, higher \lya\ EW, fainter UV magnitude, bluer UV slope, lower stellar mass, and lower SFRs show an emission-line doublet at the \civ\ rest wavelengths, with no sign of absorption on the blue side. In these cases, the nebular emission from ionized gas dominates the spectra. This is indicative of very low (Z$<0.002$) interstellar metallicity \citep[see for example Figs. 15 and 18 of][]{Vidal-Garcia2017}. We fit two Gaussian profiles to the \civ\ emission doublet, meaning under the assumption that nebular emission dominates the feature profile in the stacked spectra, without evidence of a P-Cygni profile, and we found \civ\ EWs ranging between 1.95 and 4.74 \AA. 

The \siliv\ interstellar absorption exhibits a behavior similar to that of \civ\ and appears in absorption or P$-$Cygni when \civ\ is in P$-$Cygni, while it is reduced, if not suppressed, when the \civ\ nebular emission dominates the profile. Absorption in the ISM or CGM can also play a significant role in shaping the \civ\ profile. In addition, stellar \siliv\ P$-$Cygni, which is a signature of stellar winds, can be contaminated by interstellar lines of the same element. Interestingly, the appearance of low-ionization absorption features resemble those of \civ\ and \siliv. In particular, \cii\ and \silii\ are in absorption when \civ\ shows a P-Cygni profile, as in the cases of higher \lya\ FWHM, lower \lya\ EW, redder UV slope, and more massive LAEs. This is suggestive of common variables, likely stellar mass and SFR, driving the spectral differences between the LAE subsamples, as discussed at the end of Sect. \ref{sec:dependence}.  

%-----------------------------
% Figure 6

\section{Discussion}\label{sec:discussion}

\subsection{Dependence of UV line features on galaxy properties}\label{sec:dependence}

The main goal of this paper is to investigate how the UV spectral features of continuum faint ($-20 \lesssim {\rm M_{UV}} \lesssim -16$ ) and low-mass ($10^{7} \lesssim {\rm M_{\star}} \lesssim 10^{10} \, {\rm M_{\sun}}$ ) LAEs at $2.9 \leq z \leq 4.6$ vary with their properties. These variations can be appreciated in Figs. \ref{Fig5} and \ref{Fig6} and are discussed below. \\

\subsubsection*{{\it Redshift}} The stacked spectra of LAEs in the two redshift intervals (Fig. \ref{FigA1}) do not show strong qualitative differences in the absorption and emission features. A more quantitative comparison is hampered by the low S/N at $z>3.6$. For example, the non-detections of \siliiib\ and the upper limits of \ciii\ in the higher redshift subsample are hardly ascribable to a difference in terms of physical properties, but can be  associated with a higher spectral noise of $z>3.6$ LAEs. This is because of the higher luminosity distance and the higher noise level due to the sky emission at a longer wavelength. However, the limits derived from the higher $z$ subsample are still consistent with those of the lower $z$ stacked spectra (panel (a) of Fig. \ref{Fig5}).

\subsubsection*{{\it Ly$\alpha$ luminosity and flux} }
Even though most of the emission lines are clearly visible in the mean spectra of LAEs in different bins of \lya\ luminosity and flux, the \heii, \oiiir\ and \ciiir\ lines are detected with S/N $>2.5$ only in the composite spectra of the brightest (log$_{10}$(L$_{{\rm Ly}\alpha} / ({\rm erg \, s^{-1}})) > 42.05$ and log$_{10}$(F$_{{\rm Ly}\alpha} /({\rm erg s^{-1} cm^{-2} })) > -17.02$) LAEs (panels (b) and (c) of Fig. \ref{Fig5}). This is mainly due to the, on average, higher continuum S/N of a single spectrum for the brightest sources, as most of the them have stronger \lya\ emission. This is also the case for the stacked spectra of the higher SFR bin, reflecting that, to first order, the \lya\ luminosity scales with the SFR \citep[e.g.,][]{Sobral2019b}.
The spectral stacks of the brightest LAEs (Figs. \ref{Fig3} and \ref{FigA2}) exhibit typical absorption profiles of stellar winds, such as \civ\ and \siliv. Theoretical models predict the strengths of these features to be higher at younger ages and increase from low to high stellar metallicities \citep[e.g.,][]{Vidal-Garcia2017, Byler2018}. Given the absorption features (\silii, \oisilii, \cii) in the spectra of brighter \lya\ subsamples, one cannot neglect a contribution from gas and dust in the ISM and CGM in shaping the P$-$Cygni profiles. If this is the case, the differences between the \lya-faint and -bright subsamples (Fig. \ref{Fig6}) can be explained in terms of photon scattering through a larger amount of gas and dust in the ISM of the brightest sources, which are, in general, more massive. This scenario is strengthened by the stacked spectra for LAEs grouped in mass, where the more massive LAEs show stronger P$-$Cygni profiles and absorption features. An alternative or additional explanation is that \civ\ photons are scattered out by outflowing gas in the CGM escaping at larger radii. The stacked spectrum of less luminous LAEs show \civ\ nebular emission (Fig. \ref{Fig6}), with no evidence of absorption, which is indicative of low metallicities. However, the low S/N of the stacked spectra of UV fainter sources reduces our ability to clean the nebular emission of contamination from stellar and interstellar absorption. In a recent work on z$\sim5$ star-forming galaxies, \cite{Pahl2020} concluded that the increase of \lya\ strength and the detection of strong nebular \civ\ emission points toward elevated ionized photon production efficiency. In the case of our sample, this is further supported by detection of the \heii\ recombination line (which requires photons with energy $> 54$ eV) in the stacked spectra of our brightest LAEs.  

\subsubsection*{{\it Ly$\alpha$ FWHM and EW}}
The EWs of the \ciiib\ and \oiiir\ collisionally excited emission lines are more than 0.2 dex larger in the mean spectra of LAEs with FWHM $\leq 271$ km s$^{-1}$ and with \lya\ EW $> 97.2 \AA$ (panels (d) and (e) of Fig. \ref{Fig5}). An increase of \ciii\ with increasing \lya\ EW has been observed in previous works \citep[e.g.,][]{Shapley2003, Stark2014, Nakajima2018b, Du2018}. This is interpreted as a common mechanism dominating the emission of collisionally excited UV lines and the production and escape of \lya\ radiation. A higher ionizing photon production by young and metal-poor stars for stronger \lya\ EWs reduces the neutral covering fraction and allows more photons to escape the galaxy \citep[e.g.,][]{Du2018}. This would be consistent with the scenario in which the extended \lya\ radiation \citep[more than 50\% of the total \lya\ radiation in a galaxy comes from extended haloes, e.g.,][]{Leclercq2017}, or a fraction of it, is radiation produced in star-forming regions scattered in an outflowing medium. 
However, while the relation between \lya\ and \ciii\ EWs seems to hold for LAEs with similarly high \lya\ EW to those considered in this work, it becomes weaker when galaxies with lower \lya\ EW or \lya\ in absorption are considered. For example, \citet[][Fig. 6]{LeFevre2019} found a broad scatter between the EW of \lya\ and that of \ciii, and also a fraction of galaxies with significant \ciii\ emission and \lya\ in absorption.

For LAEs with higher FWHM or smaller EW, the \ciiib\ and \oiiir\ EWs are smaller because of the stronger underlying continuum as result of a more intense star formation. The mean SFR is 0.3 and 1.2 M$_{\odot}$ yr$^{-1}$ for lower and higher FWHM subsamples splits, respectively. Similarly, the LAEs have a mean SFR of 0.4 and 1.35 M$_{\odot}$ yr$^{-1}$ for the larger and smaller EW subsamples. 

\civ\ is in emission for the LAEs with lower FWHMs, while in the case of higher FWHMs, the \civ\ P-Cygni profile and interstellar absorption features are clearly visible (Fig. \ref{Fig6}). The \civ\ doublet is observed in emission (\civ\ EW of 4.2\AA) for large \lya\ EW, while the \civ\ profile is unclear for the stacked spectra of the subsample with smaller \lya\ EW. If we neglect the stellar continuum, these differences in the \lya\ and \civ\ profiles between the subsamples with lower/higher FWHMs and larger/smaller EWs can be interpreted as radiation transfer through a different amount of gas. The emission from resonant lines can be similarly affected if they are seen through the same gas, as is also shown for the \lya\ and \mgii $\lambda2800$ doublet in local ``green pea'' galaxies \citep[e.g.,][]{Henry2018}. However, \lya\ and \civ\ trace different phases of the medium within and around galaxies. While \lya\ profiles are shaped by resonant scattering in the neutral gas, \civ\ traces the highly ionized gas. Our result would suggest that a lower amount of neutral gas implies a lower amount of ionized gas. A Spearman's correlation coefficient of 0.52 supports a moderate dependence of \lya\ FWHM with stellar mass. Similarly, smaller EW LAEs (which would correspond to a lower amount of gas) have higher stellar masses. This is consistent with the mean spectra of LAEs with higher stellar mass exhibiting \civ\ in P-Cygni profile. 
A higher fraction of neutral gas for higher \lya\ FWHM and lower \lya\ EWs is also supported by the presence of stronger low-ionization absorption features in the stacked spectra of these subsamples. A correlation between low-ionization absorption lines and \lya\ EW, stellar mass, UV luminosity, and $\beta$ slope has been observed in LBG at $z \approx 3$ by \cite{Jones2012} and explained in terms of star formation-driven outflows of neutral gas responsible for \lya\ scattering and the strong low-ionization absorption lines, while increasing stellar mass, metallicity, and dust content. Our results are consistent with their picture, as low-ionization absorption features are stronger for redder UV slopes (higher dust content), larger stellar mass, higher SFR, and brighter LAEs.

\subsubsection*{{\it UV magnitude and spectral slope}}
The stacked spectra of the UV fainter LAEs exhibits larger ($> 0.5$ dex) \oiiir\ and \ciiib\ EWs than the UV brighter subsample (panel (f) of Fig. \ref{Fig5}). 
Very interestingly, a clear \civ\ in emission (\civ\ EW of 3.9 \AA) is observed in the mean spectra of the fainter (M$_{\rm UV} >-17.9$) LAEs (Fig. \ref{Fig6}), indicative of a hard ionization field, such as that from young and massive stars, and of an elevated ionized photon production efficiency. 
The UV spectral slope is considered a proxy for the dust content as well as for the hardness of the ionizing radiation. Since the fainter sources have, on average, bluer colors (Fig. 2 of H17), the subsample splits in $\beta$ present similar stacked spectra to those of the M$_{\rm UV}$ subsamples (Fig. \ref{FigA5} and \ref{FigA6}). With the cautiousness on the uncertainties related to the $\beta$ slope calculation (Sect. \ref{sec:properties}) in mind, we note that LAEs with bluer UV slopes show \civ\ in emission (\civ\ EW of 1.95 \AA). They also have \ciii\  EWs more than 0.2 dex larger than LAEs with higher $\beta$. In the case of LAEs with redder slopes, deep absorption features suggest the presence of a higher fraction of neutral gas in the ISM and CGM of these sources. 
The harder ionization field of the bluer LAEs is confirmed by the detection of the \heii\ line in emission. The UV fainter LAEs of our sample have, on average, higher \lya\ emission strengths (LAEs with M$_{\rm UV} > $ and $\leq -17.9$ have average \lya\ EWs of 201 and 76 \AA, respectively), which implies higher ionzing photon production rate in stronger LAEs. 
These results are in line with current detections of \civ\ and \heii\ in emission in low-mass, UV-faint star-forming systems in the local Universe \citep[e.g.,][]{Senchyna2017, Berg2019} and at $z\approx 2-3$ \citep[e.g.,][]{Berg2018, Nakajima2018b}. \\

\subsubsection*{{\it Stellar mass, SFR, and sSFR}}
The stacked spectrum for the lower stellar mass bin shows, in addition to the \civ\ doublet in emission, \ciiib\ and \oiiir\ EWs are 0.3 dex larger compared to that of the higher mass bin (panel (h) of Fig. \ref{Fig5}). At the same time, the stacked spectra of the higher mass LAEs show absorption features that are not clearly observed in the spectra of the lower mass subsample (Fig. \ref{FigA7}). The \ciiib\ and \oiiir\ EWs are stronger (0.6 and 0.4 dex, respectively, panel (i) of Fig. \ref{Fig5}) for the lower SFR subsample as LAEs with low SFR are, on average, less massive. High SFR implies a strong emission continuum. 
As already mentioned, the \civ\ doublet varies from emission to P$-$Cygni with increasing stellar mass and SFR. A similar behavior has been observed for the also resonant \mgii\ doublet \citep[e.g.,][]{Finley2017, Feltre2018}. 

The mean spectra of LAEs with higher sSFR have larger \ciii\ EW. This has been found to correlate with the EW of the optical \oiiibpt\ line \citep{Maseda2017}, which in turn correlates with the sSFR for low-metallicity starbursts \citep{Tang2019}. It is not possible to discuss the trends of the other features as a function of sSFR because of the difficulty in detecting emission features in the stacked spectra of LAEs with higher sSFR, as these are also among those with the lowest UV luminosity, and therefore have individual spectra of lower S/N. 

The similarity of the variation of spectral features among the LAE subsamples implies that these spectral differences contain important information on the physics of galaxies and that these are mainly dictated by SFR and stellar mass, which are intrinsically linked to differences in ages, metallicity and dust content of galaxies.  \\

\subsection{Comparison with the literature}\label{sec:comparison}

\begin{figure*}
  \centering
   \includegraphics[width=14.5cm]{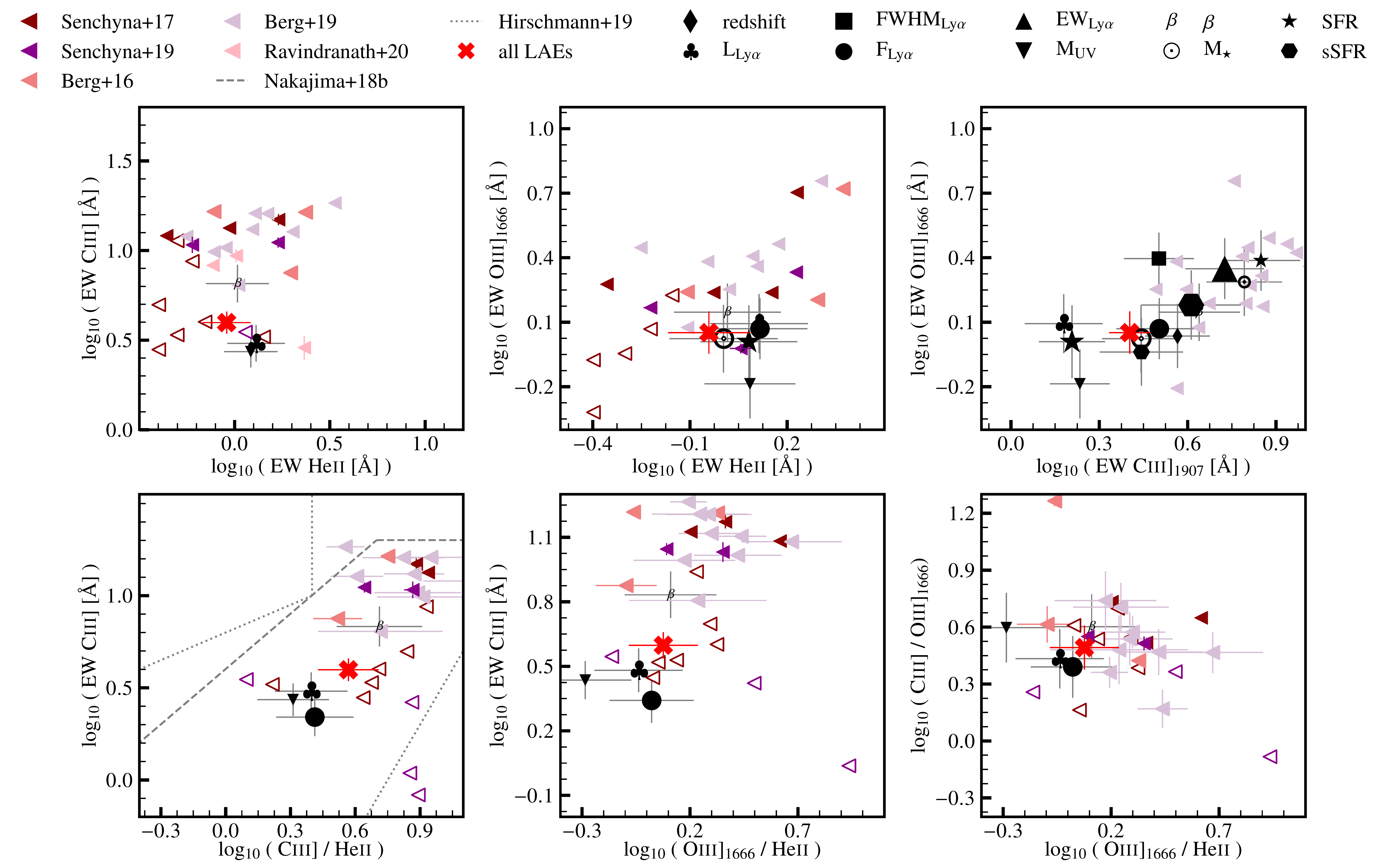}
   \caption{UV emission line properties of MUSE HUDF LAE subsamples and local galaxies from the literature described in Sects.~\ref{sec:nebular_lines} and \ref{sec:comparison}. The diagrams show different combinations of EWs and ratios of the \heii, \oiiir, \ciii,\  and \ciiib\ emission lines. Different black symbols refer to S/N $> 2.5$ line detections for different LAE subsamples and the red cross to the total LAE sample. Smaller and larger symbols indicate the subsample of LAEs whose property is lower or higher than the median value, respectively.  Different colors refer to different samples from the literature, as labelled in the legend. Open and filled symbols indicate upper limits and detections, respectively. The left triangles refer to local galaxies. The gray dashed lines are the separation criteria between AGN and star-forming galaxies from \cite{Nakajima2018b}, while the gray dotted lines represent the selection criteria for the AGN, composite, and star-forming galaxies from \cite{Hirschmann2019}.}
            \label{Fig7}%
   \end{figure*}

\begin{figure*}
  \centering
   \includegraphics[width=14.5cm]{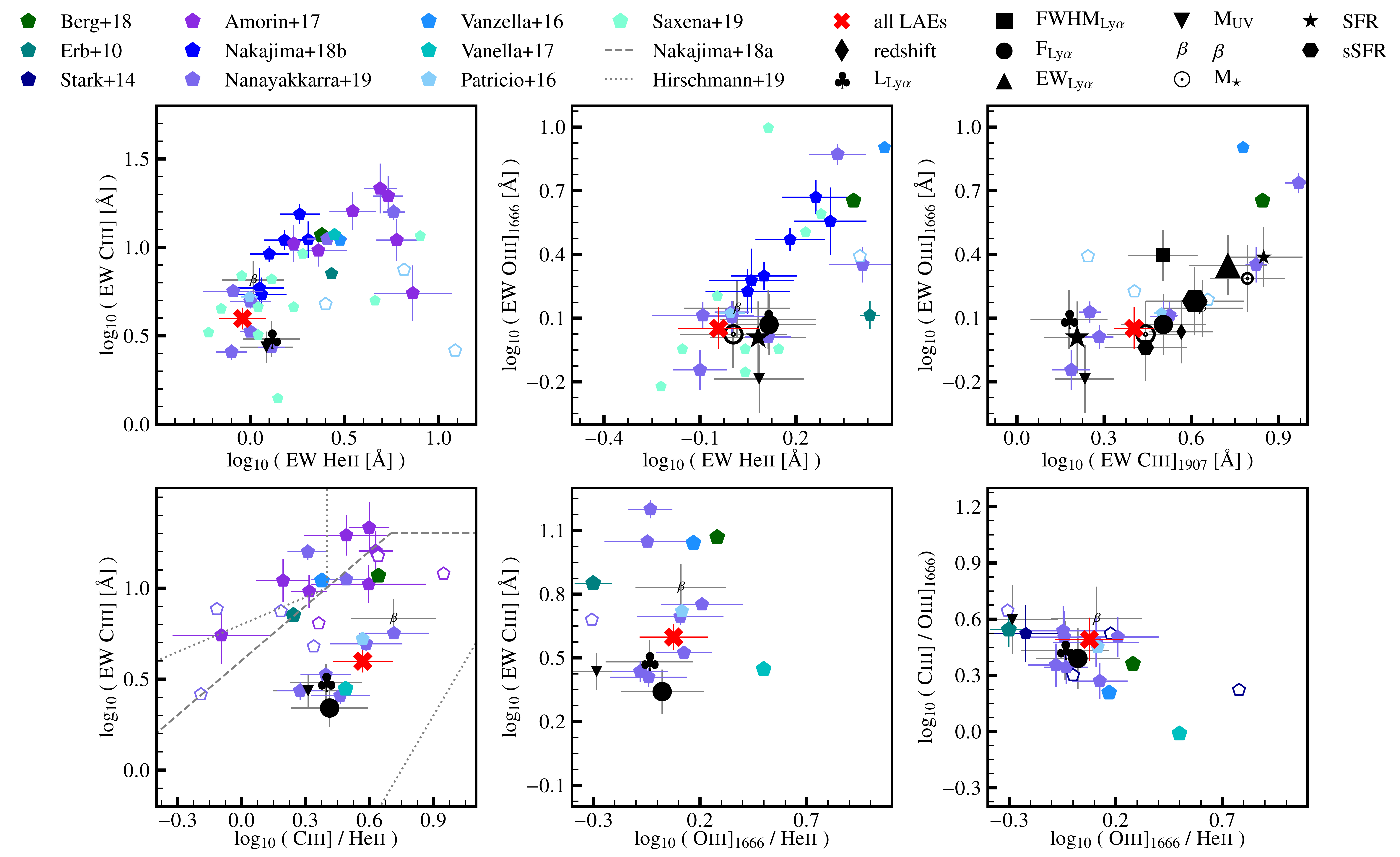}
   \caption{Same as Fig. \ref{Fig8}. Open and filled symbols indicate upper limits and detections, respectively. The pentagons refer to data for $2 \leq z \leq 4$ galaxies.}
            \label{Fig8}%
   \end{figure*}

With caution for uncertainties on the systemic redshift in mind (Sect. \ref{sec:lya_correction}), we investigated how the line measurements described in Sect.~\ref{sec:nebular_lines} compare with those already presented in the literature. 
We considered data from local metal-poor galaxies (Fig. \ref{Fig7}) and strong line emitters at $z\geq2-3$ (Fig. \ref{Fig8}), which are both considered valuable examples of the young galaxies that could have significantly contributed to the ionizing photon budget necessary to sustain cosmic reionization. 
Figures \ref{Fig7} and \ref{Fig8} show comparisons of EWs and line ratios of the MUSE HUDF LAEs with those from the literature. The spectra of our LAEs have not been corrected for potential attenuation by dust in order to avoid using an arbitrary attenuation curve, while the data from the literature have been corrected as described in the corresponding works. The impact of this on our analysis is negligible, as we considered the \ciii/\heii, \ciii/\oiiir, \oiiir/\heii\ line ratios that would differ by $\approx$ 0.1 dex assuming a V-band attenuation of one order of magnitude (A$_{\rm V}=$1) and the \cite{Calzetti2000} curve \citep[see also, Section 5.1 of][]{Hirschmann2019}. 
By adopting an A$_{\rm V}=1$ mag, we would vastly overestimate the dust content of our LAEs. As discussed in Sect.~\ref{sec:stack_methodology}, the SED fitting to the HST photometry indicates a mean attenuation of 0.1 mag in the $V$-band, and the blue UV slopes of our LAEs are in line with a low dust content.  \\

{\it Comparison with local metal-poor galaxies.} Figure \ref{Fig7} shows measurements performed on local low-mass, metal-poor ($Z \lesssim 10-20\% ~Z_{\odot}$) star-forming systems with sSFRs from 1 up to $\sim 10^2$ Gyr$^{-1}$ \citep{Berg2016, Berg2019, Senchyna2017, Senchyna2019} and $0.13 < z < 0.3$ green pea galaxies from \cite{Ravindranath2020}. In particular, \cite{Senchyna2017} observed a marked transition in the spectral properties of the UV features with decreasing metallicity. Their star-forming systems with $Z < 1/5 ~Z_{\odot}$ showed more prominent nebular emission in \heii\ and \civ,\ and weak, if not absent, stellar wind features compared to the less metal-poor targets in their sample. Our LAE subsamples show similar spectral variations. For example, the prominent nebular emission lines (\civ, \oiiisf, \ciii) observed in lower mass, UV faint but higher \lya\ EW LAEs require a strong ionizing radiation field and unveil the role of young and metal-poor massive stars in dominating the spectra of these sources. This reveals a strong interplay between the physical properties, such as stellar mass and SFR, and therefore age and metallicity, in driving the differences in spectral features. 

The \oiiib\ and \ciiib\ EWs of the mean spectra of our LAEs are, on average, lower (0.3 and 0.5 dex, respectively) than those in the local systems (Fig. \ref{Fig7}) and compatible only with the upper limits from \cite{Senchyna2017, Senchyna2019}. Our LAEs have emission line EWs similar to those of the local metal-poor galaxies for the bluer, lower stellar mass and higher \lya\ EW subsamples.  These are the properties that characterized the local galaxies considered for this comparison. The \ciii\ EWs of the $0.13 < z < 0.3$ green pea galaxies from \citet[][Table 3]{Ravindranath2020} are $<10 \,\AA$ and reach values as low as the lowest of our LAEs, in addition to 1-$\sigma$ upper limits with \ciii\ EW $< 1 \,\AA,$ indicating even weaker emission for the green pea galaxy sample. 

The ratios of the collisionally excited \oiiir\ and \ciii\ lines to the \heii\ recombination line is about 0.2 dex lower in our LAEs than the ratios measured in the local Universe, suggesting a more intense ionizing radiation field for our LAEs, which would increase the \heii\ emission. This is also supported by the stronger EWs of the \civ\ nebular emission measured on our stacks compared to those of the local sources. An increase from low to high $z$ in the ionization parameter (i.e., the ratio of the number of H-ionizing photons to the number of atoms of neutral hydrogen), which is linked to SFR, or in the ionizing photons' production efficiency can explain this small difference in line ratios. 
In addition, the differences in the spectral properties of local sources are affected by the target selection, given that the low-$z$ galaxies considered here have been selected to be extremely metal-poor galaxies with bright optical emission lines.\\

{\it Comparison with $z \approx 2 - 4$ galaxies.} High-ionization UV lines have been detected in the spectra of $z\gtrsim2$ galaxies through gravitational lensing \citep[e.g.,][]{Stark2014, Patricio2016, Vanzella2016, Vanzella2017, Berg2018}, spectral stacking \citep[e.g.,][]{Nakajima2018b,Rigby2018,Saxena2019}, and deep spectroscopic observations \citep[e.g.,][]{Erb2010,Maseda2017,Amorin2017,Nanayakkara2019}. The EWs of \heii, \oiiir,\ and \ciii\ from these works are larger than those measured in the average spectra of our LAEs, with the exception of some of the MUSE sources at $z\gtrsim 3$ studied by \cite{Patricio2016} and \cite{Nanayakkara2019}, some of the stacks from \cite{Nakajima2018b}, and some of the VANDELS \heii\ emitters from \cite{Saxena2019}.
The EWs from \cite{Nakajima2018b} are computed from average spectra of a z$\approx3$ LAE population whose median UV luminosity is about two orders of magnitude brighter than ours. Most of the $2.4 < z < 3.5$ UV-selected, low-luminosity galaxies from \cite{Amorin2017} have higher \heii\ and \ciii\ EWs, most likely because of their higher SFR (see Fig. \ref{Fig1b}). The line measurements  from \cite{Saxena2019} are performed on the stacked spectra of UV continuum, bright \heii\ emitters ($-19 < {\rm M_{UV}} < -22$). The strong line emitters from \cite{Erb2010} and \cite{Berg2018} are brighter and more massive than the median value of our LAEs. The lensed galaxies from \cite{Vanzella2016,Vanzella2017} are among the least massive, most metal-poor, young, and faintest systems observed at $z\sim 3$. With M$_{\rm UV}> -16$, the source ID14 from \cite{Vanzella2017} is roughly one order of magnitude fainter that the faintest LAE in our sample, while the source ID11 from \cite{Vanzella2016} has an remarkable blue UV slope ($\beta = -2.95$). 
Recently, \cite{Du2020} measured \ciii\ EWs in $z\sim2$ analogs of galaxies in the reionization era, obtaining values from $13.2 \, \AA$ down to $1 \, \AA$, reaching values as low as those of our LAEs. The authors also found differences in the \ciii\ EW depending on the selection criteria, with higher values of \ciii\ EW for emission lines rather than for continuum-selected galaxies.
\cite{Mainali2020} detected two targets with \ciii\ emission reaching EW $\approx 17 - 21 \, \AA$ in the spectra of $z\sim2$ galaxies selected for their strong rest-optical line (\oiiibpt$+$H$\beta$) EWs. These values are above the ones of our LAEs and are similar to those observed at $z>6$  \citep[e.g.,][]{Stark2015a, Stark2017, Hutchison2019}.

Interestingly, the line ratios (bottom panels of Fig. \ref{Fig8}) of  $z \approx 2 - 4$ galaxies from the literature are consistent with those measured for our LAEs, suggesting comparable ISM conditions at these similar redshift ranges. The fact that the targets from the aforementioned published works have been selected as strong line emitters can explain the difference in terms of EW strengths. Our results indicate that these extreme emitters may not be representative of the whole population of UV-faint, low-mass LAEs at $z\gtrsim3$.

{\it Comparison with $z \gtrsim 6$ galaxies.} In recent years, the detection of high-ionization nebular emission lines, such as \civ, \ciii,\ and \heii, has been possible in the deep rest-frame UV spectra of $z\gtrsim6$ rare, very bright, or gravitationally lensed galaxies \citep[e.g.,][]{Sobral2015, Sobral2019a, Stark2015a, Stark2015b, Mainali2017, Schmidt2017, Mainali2018, Hutchison2019}. In particular, the \civ\ EWs measured from the spectra of high-$z$ galaxies with UV magnitude in the same range as our LAEs (M$_{\rm UV} \geq -21$) are larger than 20 \AA\ \citep[see figure 6 of][]{Mainali2018}. These extreme values are compatible with those measured in obscured AGN but also with photo-ionization by young and massive stars \citep{Stark2015b}.
These values are, however, at least four times higher than the \civ\ EWs of the average of our LAE subsamples showing the \civ\ doublet in emission. This could reflect a harder ionizing radiation field associated with young, hot, and metal-poor stars, as well as a much higher ionizing photon production efficiency at high redshift \citep[Sect. 4.4 of][]{Nanayakkara2020}. We also have not yet detected a less extreme population of $z \gtrsim 6$ faint and low mass galaxies exhibiting spectral properties similar to the average ones of faint LAEs at $3 \leq z \leq 4.6$, probably because of the selection methods \citep[e.g.,][]{Du2020}. Spectroscopic studies of galaxies in the first billion years ($z\gtrsim 6$) are still confined mainly to a few gravitationally lensed or exceptionally bright sources with spectral information limited to a few emission lines, along with upper limits, per target, preventing any statistical comparison with lower $z$ sources.

%-----------------------------
% Figure 8
%

%-----------------------------
% Figure 8

\subsection{AGN and other ionizing sources}\label{sec:agn_cont}

It is worth investigating the possible contributions from AGN to the average spectra of the MUSE HUDF LAEs. 
We can exclude the possibility that the mean spectra of our LAEs are dominated by AGN for several reasons. First, our LAEs do not have \lya\ FWHMs larger than 1000 km s$^{-1}$, which would indicate a broad-line AGN \citep[see e.g.,][]{Sobral2017}. Second, the LAE subsample with the larger FWHM does not show the \civ\ doublet, which can be used as proxy for AGN activity \citep[e.g.,][]{Mignoli2019}, and we do not detect \nv\ emission, which requires photons with energy $> 77.4$ eV and is usually a tracer of the presence of an AGN \citep[e.g.,][]{Laporte2017,Sobral2018}. In addition, we discarded the objects defined as X-ray AGN in the 7 Ms Source Catalogs of the Chandra Deep Field-South Survey \citep[][see Section 4.5 for source classification]{Luo2017} from our sample. Even the deepest X-ray stacks yielded no signal for high-$z$ LAEs \citep{Urrutia2019, Calhau2019}. 

Moreover, by exploring diagnostic diagrams based on EWs and ratios of other rest-UV emission lines \citep[\civ, \heii, \oiiisf, \ciii, e.g.,][]{Feltre2016, Nakajima2018a, Hirschmann2019}, we note that the \ciii/\heii\ and \oiiir/\heii\ ratios of our LAEs are $> 1$ and 0.3, respectively, and thus not compatible with pure AGN photo-ionization models \citep[figure A1 of][]{Feltre2016}. The low values of \ciii\ and \civ\ EW ($\leq \, 20 \AA$) exclude a dominant AGN contribution to the spectral emission \citep{Nakajima2018a, Plat2019}. Our LAEs would instead be classified as composite (AGN and star-forming) galaxies in the diagrams from \cite{Hirschmann2019} whose selection criteria are, however, customized for massive galaxies ($\gtrsim 10^{9.5}$ M$_{\odot}$). 
We caution that current photo-ionization models for star-forming regions can underestimate, even by one order of magnitude, the \heii\ nebular emission of local metal-poorlocal met (below 1/5 solar), dwarf, and $z\gtrsim2$ galaxies  \citep[e.g.,][]{Jaskot2016, Steidel2016, Senchyna2017, Berg2018}, including those detected in our stacks, and MUSE individual spectra  \citep[e.g.,][]{Nanayakkara2019}. 
In this context, the reliability of the UV selection criteria for AGN for our LAEs may be reduced, and further improvements in the modeling of the \heii\ ionizing flux from young and metal-poor stellar populations (single stars or binaries) could bring the predictions for nebular emission from star-forming galaxies into agreement with the spectral properties measured in our LAEs.  

The presence of strong \heii\ emission implies the need for a hard ionizing spectrum \citep[see e.g., Fig. 1 of][]{Feltre2016} able to provide photons with sufficient energy to doubly ionize helium (54 eV) that, subsequently, recombines giving rise to \heii\ emission. At the moment, in addition to refinements to the models of young and metal-poor stars, an additional source of harder ionizing radiation seems to be the most likely explanation to account for the measured \heii\ flux. This includes weak AGN or a contribution from other sources, such as exotic Pop III stars, X-ray binaries, radiative shocks, and super-soft X-ray sources such as accreting white dwarfs \citep{Shirazi2012, Kehrig2015, Woods2016, Schaerer2019, Plat2019}. 

\section{Summary and conclusions}\label{sec:conclusions}

The MUSE HUDF enabled spectroscopic observations of faint ($-20 \lesssim {\rm M_{UV}} \lesssim -16$) and low-mass (between $ 10^7$ and $10^{10} \, {\rm M_{\odot}} $) LAEs at $z\gtrsim 3$. Information about the gas emission in these systems comes mainly through the \lya\ line which is the strongest emission line in UV spectra. We selected a sample of 220 LAEs and performed spectral stacking to explore the UV emission and absorption features that are too weak to be detected in single spectra. Stacked spectra were computed for different subsamples, partitioned on the basis of both observed (\lya\ luminosity, flux, FWHM and EW, UV magnitude, and UV slope) and physical (stellar mass, SFR, and sSFR) properties of the LAEs. 

The main focus of this work was to investigate spectral differences of the ionized gas emission features for LAEs with different properties. We were able to detect emission lines such as \heii, \oiiisf, \siliii,\ and \ciii, as well as the \civ\ doublet.
The main results of this analysis are summarized below. 
\begin{itemize}
\item With the obvious exception of \lya,\, the individual deep MUSE spectra do not show the presence of other UV lines, which are clearly detected in the spectra of galaxies. That they appear in the stacked spectra suggests weak UV emission is ubiquitous in the general population of faint and low-mass LAEs, regardless of the differences among the LAE stacks (Sects.~\ref{sec:nebular_lines} and \ref{sec:civ}). 

\item The \ciii\ and \oiiisf\ collisional excitation doublets vary with the observed properties in which LAEs have been subsampled (Sect.~\ref{sec:dependence}). In particular, we find that the \ciii\ EW of our $2.9 \leq z \leq 4.6$ LAEs increases with that of \lya\ EW, similarly to what observed at $z\approx2$ in faint dwarf galaxies \citep{Stark2014} and in stacked spectra of $z\approx3$ galaxies that are on average much brighter than our LAEs \citep{Shapley2003, Nakajima2018b}. We note, however, that the relation between the \lya\ and \ciii\ EW can present a larger scatter when extended to LAEs with lower \lya\ EWs and \lya\ emission in absorption \citep[e.g.,][]{LeFevre2019}. We find larger \ciii\ and \oiiisf\ EWs in the spectral stacks of the UV fainter and bluer LAEs.

\item We detect the \heii\ emission feature in the mean spectra of brighter, bluer and intensively star-forming LAEs of our sample (Sect. \ref{sec:nebular_lines}). The strengths of \heii\ emission are consistent with those observed in other local and $z>2$ galaxies, which are challenging current stellar evolution models (Sect. \ref{sec:agn_cont}).  
\item We find two different main profiles for the \civ\ doublet, namely P-Cygni profiles and emission, without clear signs of blue-shifted absorption. These different \civ\ profiles encode information on the main mechanisms shaping this feature, such as stellar winds from massive stars, nebular emission from ionized gas, and the presence of a neutral medium and outflowing gas. At the same time, the differences that we observe among the stacked spectra suggest that the shape of the \civ\ profile could also be used as a proxy for galaxy properties. For example, we observe  \civ\ purely in emission in the stacked spectra of low stellar mass, faint, and blue LAEs at $2.9 \leq z \leq 4.6$ (Sects.~\ref{sec:civ} and \ref{sec:dependence}). 

\item The emission features detected in our LAE stacked spectra are overall in agreement with those measured for other MUSE sources at similar redshifts and with metal-poor, low-mass star-forming systems in the local Universe (Sect.~\ref{sec:comparison}). One exception is the \heii\ and \ciii\ EWs of $z\sim2$ analogs of galaxies in the reionization era, which are higher than the average values of our LAEs, likely because of the emission-line selection criteria and their intense SFR, or both. Moreover, the \oiiir/\heii\ and \ciii/\heii\ ratios of our LAEs are smaller than those of the local dwarf galaxies. This is likely to be because of a variation in production efficiency,  strength, and hardness of the ionizing radiation. 
\end{itemize}

The variations in the emission and absorption features in the different spectra are mainly dictated by SFR and stellar mass, which are intimately related to the stellar ages, metal, and dust content of galaxies. This observational evidence could be coupled with theoretical predictions for galaxy evolution to investigate whether the observed trends of the spectral features enable the identification of a given galaxy's evolutionary phase in the SFR-M$_{\star}$ plane.

 The stacked spectra presented in this work can be exploited to understand the properties of stellar populations and ionized-versus-neutral gas in faint, low-mass LAEs at high $z$. While the presence of carbon and oxygen UV lines (\oiiisf, \ciii,\ and \civ) provides important constraints on the C/O abundance ratio, the absence of hydrogen lines other than \lya\ in the UV regime, whose complex resonant nature seriously hampers any estimate of the intrinsic nebular flux, may challenge the estimates of the oxygen abundance (O/H). The latter can be directly probed through rest-frame optical spectroscopy (e.g., \oiiibpt\ and hydrogen Balmer lines), which, in addition, offers the possibility of exploiting standard strong line diagnostics of the properties of the ionized gas such as metallicity, density and ionization level. Regarding our LAEs, additional information from rest-frame optical lines (e.g., from Keck/MOSFIRE or, in the future, from JWST/NIRSpec) will play a crucial role in determining the physical conditions (SFRs, ionization conditions, and gas-phase metallicities) within them, and further constraining the ionizing photon production rates from stellar population models. 
 
 To conclude, our stacked spectra are the only available empirical templates of faint and low-mass LAEs, and they will be instrumental to the design of spectroscopic observations of higher $z$ galaxies, including targets in the epoch of reionization, with future facilities (e.g., JWST, ELT). 
 The average spectra computed in this work are available electronically \footnote{\url{http://muse-vlt.eu/science/data-releases/} and \url{http://cdsarc.u-strasbg.fr}}.

\begin{acknowledgements} 
We thank Margherita Talia, St\'ephane Charlot, Adele Plat and Alba Vidal-Garc\'ia for helpful discussions. This work is supported  by the ERC advanced grant 339659-MUSICOS (R.  Bacon). AF acknowledges the support from grant PRIN MIUR 2017 20173ML3WW. MVM and JP would like to thank the Leiden/ESA Astrophysics Program for Summer Students (LEAPS) for funding at the outset of this project. FL, HK, and AV acknowledge support from the ERC starting grant ERC-757258-TRIPLE. TH was supported by Leading Initiative for Excellent Young Researchers, MEXT, Japan. JB acknowledges support by FCT/MCTES through national funds by the grant UID/FIS/04434/2019, UIDB/04434/2020 and UIDP/04434/2020 and through the Investigador FCT Contract No. IF/01654/2014/CP1215/CT0003. HI acknowledges support from JSPS KAKENHI Grant Number JP19K23462. We would also like to thank the organizers and participants of the Leiden Lorentz Center workshop: Revolutionary Spectroscopy of Today as a Springboard to Webb.
This work made use of several open source python packages: \textsc{NUMPY} \citep{vanderWalt2011}, \textsc{MATPLOTLIB} \citep{Hunter2007}, \textsc{ASTROPY} \citep{Astropy2013} and \textsc{MPDAF} \citep[MUSE Python Data Analysis Framework,][]{Piqueras2019}.  

\end{acknowledgements}

% for the bibliography, at the end
\bibliographystyle{aa} % style aa.bst
\bibliography{aa_ms} % your references Yourfile.bib

\begin{appendix}
\section{Average spectra of LAEs for the whole sample and subsamples}\label{app:A}

%-----------------------------
 % Figure A1
 \begin{figure*}
 \centering
  \includegraphics[width=16.5cm]{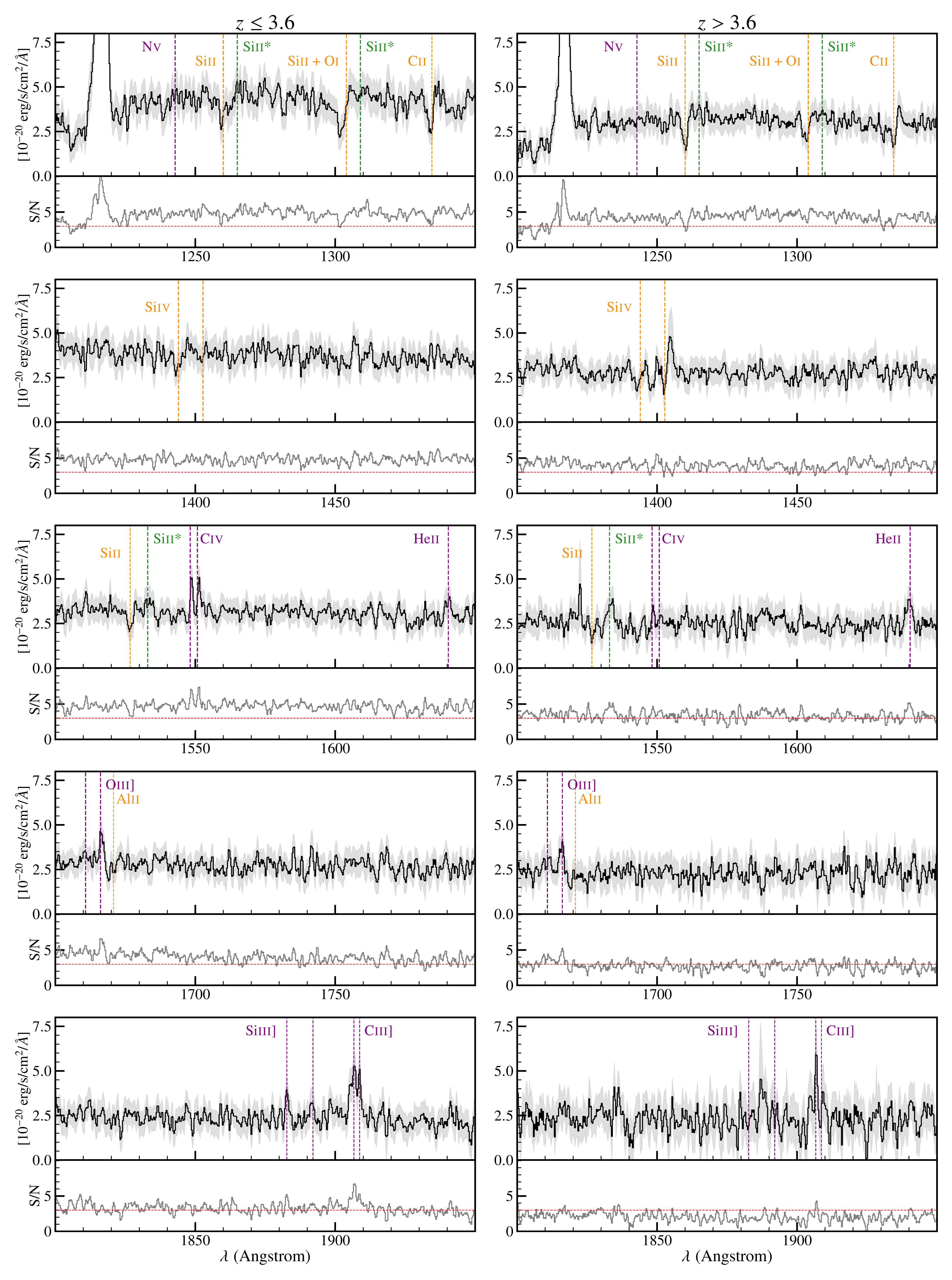}
  \caption{Mean spectra of LAEs with $z\leq3.6$ and $>3.6$ (left and right, respectively). The median values of z for the two subsamples are 3.3 and 4.0, respectively. Lines and symbols as in Fig. \ref{Fig3}.}
           \label{FigA1}%
   \end{figure*}

% Figure A2
 \begin{figure*}
  \centering
   \includegraphics[width=16.5cm]{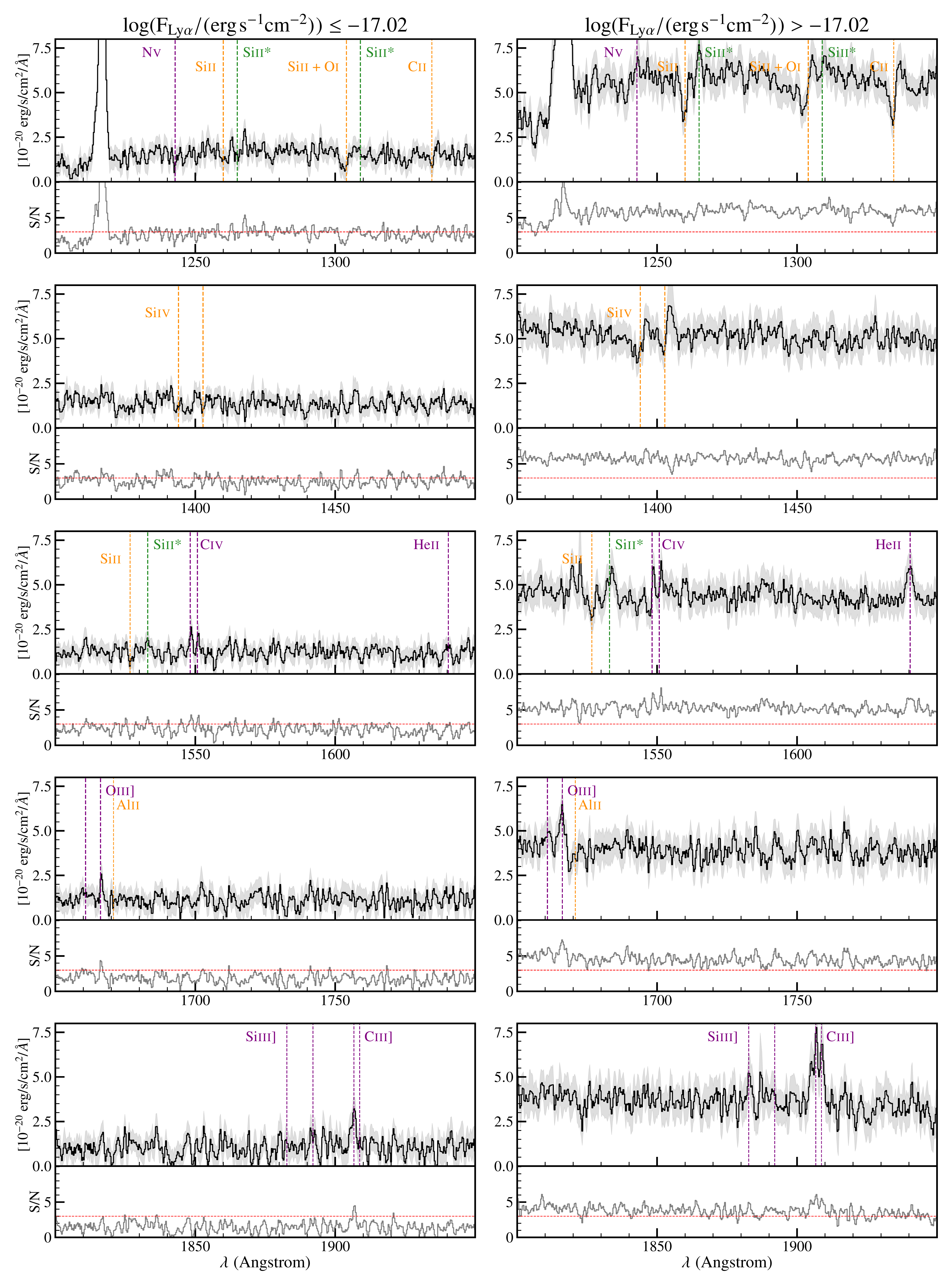}
   \caption{Mean spectra of LAEs with log$_{10}$(F$_{{\rm Ly}\alpha} / {\rm (erg \, s^{-1} cm^{-2}})) \leq -17.02$ and $> -17.02$ (left and right, respectively). The median values of log$_{10}$(F$_{{\rm Ly}\alpha} / {\rm (erg \, s^{-1} cm^{-2}}$)) for the two subsamples are $-$17.24 and $-$16.78, respectively. Lines and symbols as in Fig. \ref{Fig3}.}
            \label{FigA2}%
   \end{figure*}
%

% Figure A3
  \begin{figure*}
  \centering
   \includegraphics[width=16.5cm]{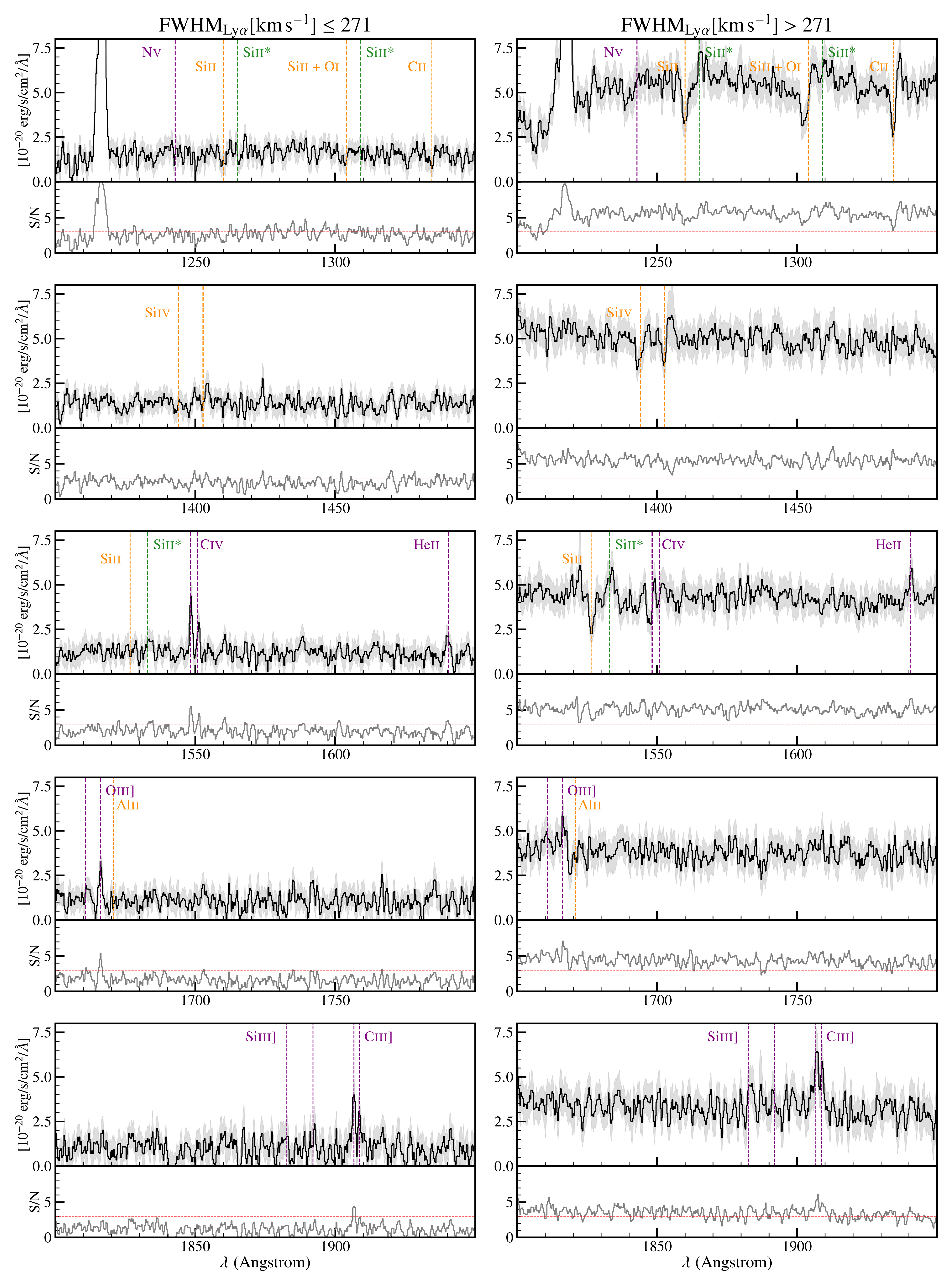}
   \caption{Mean spectra of LAEs with \lya\ FWHM $\leq 271$  and $> 271$ km s$^{-1}$ (left and right, respectively). The median values of the Ly$\alpha$ FWHM for the two subsamples are 222 and 360, respectively. Lines and symbols as in Fig. \ref{Fig3}.}
            \label{FigA3}%
   \end{figure*}

% Figure A4
 \begin{figure*}
  \centering
   \includegraphics[width=16.5cm]{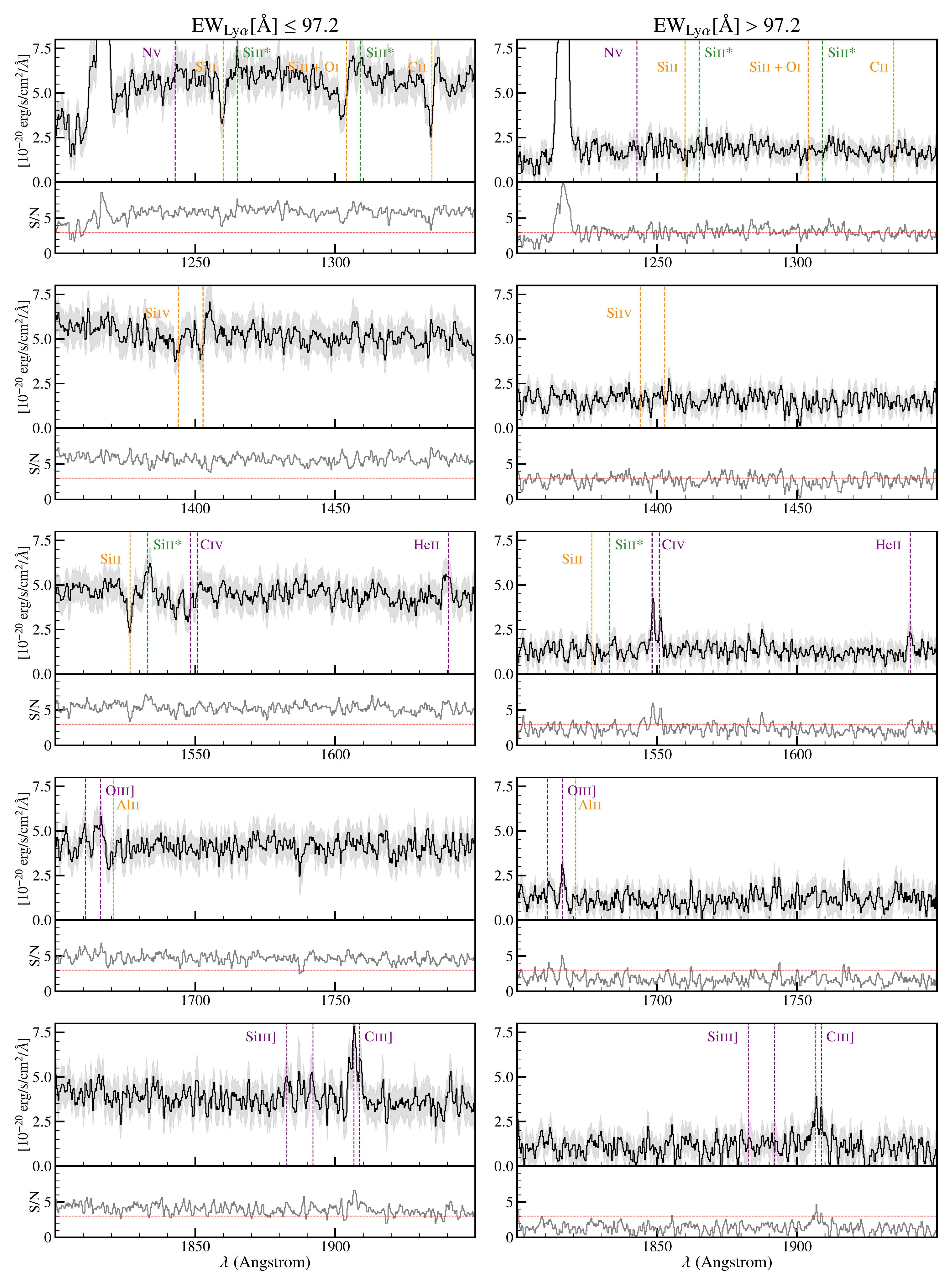}
   \caption{Mean spectra of LAEs with \lya\ EW [\AA] $\leq 97.2$  and $> 97.2$ (left and right, respectively).  The median values of the Ly$\alpha$ EW for the two subsamples are 54.7 and 163.1, respectively. Lines and symbols as in Fig. \ref{Fig3}.}
            \label{FigA4}%
   \end{figure*}
%

% Figure A5
  \begin{figure*}
  \centering
   \includegraphics[width=16.5cm]{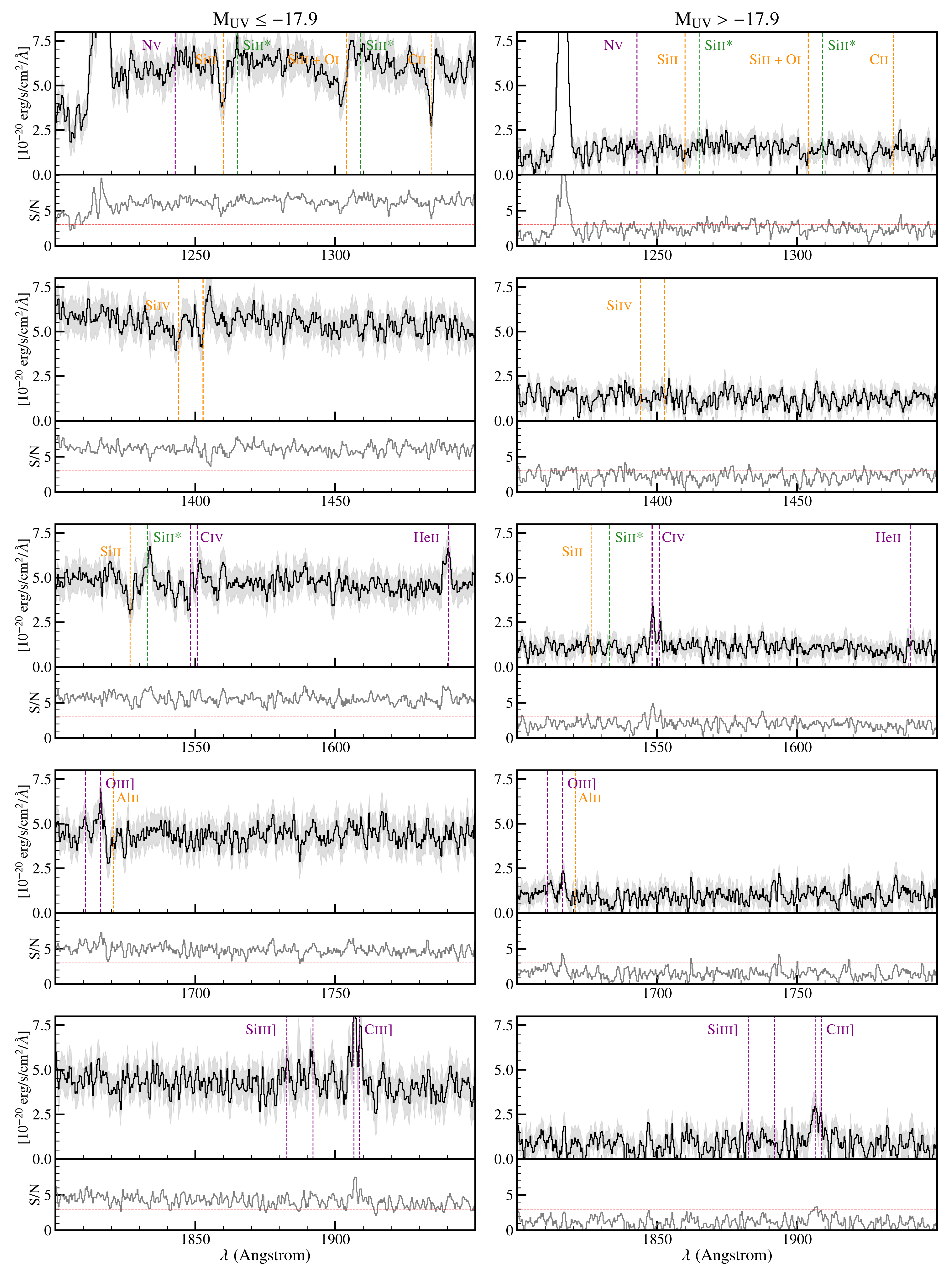}
   \caption{Mean spectra of LAEs with M$_{\rm UV} \leq-17.9$  and $> -17.9$ (left and right, respectively). The median values of the M$_{\rm UV}$ for the two subsamples are $-18.8$ and $-17.2$, respectively. Lines and symbols as in Fig. \ref{Fig3}.}
            \label{FigA5}%
   \end{figure*}
%

% Figure A6
  \begin{figure*}
  \centering
   \includegraphics[width=16.5cm]{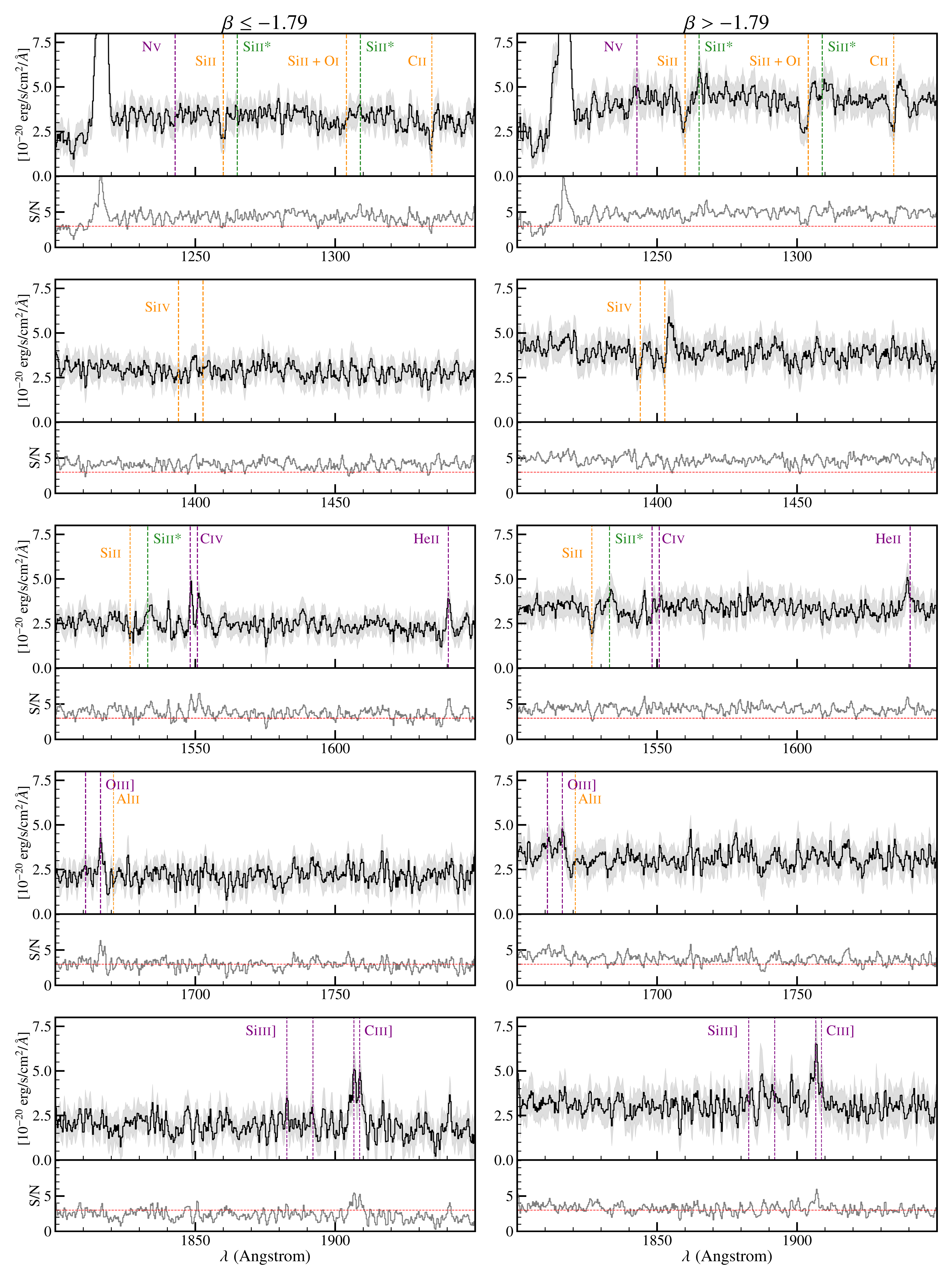}
   \caption{Mean spectra of LAEs with $\beta \leq-1.78$ and $> -1.78$ (left and right, respectively). The median values of  $\beta$ for the two subsamples are $-2.07$ and $-1.40$, respectively.  Lines and symbols as in Fig. \ref{Fig3}.}
            \label{FigA6}%
   \end{figure*}

 %Figure A7
  \begin{figure*}
  \centering
   \includegraphics[width=16.5cm]{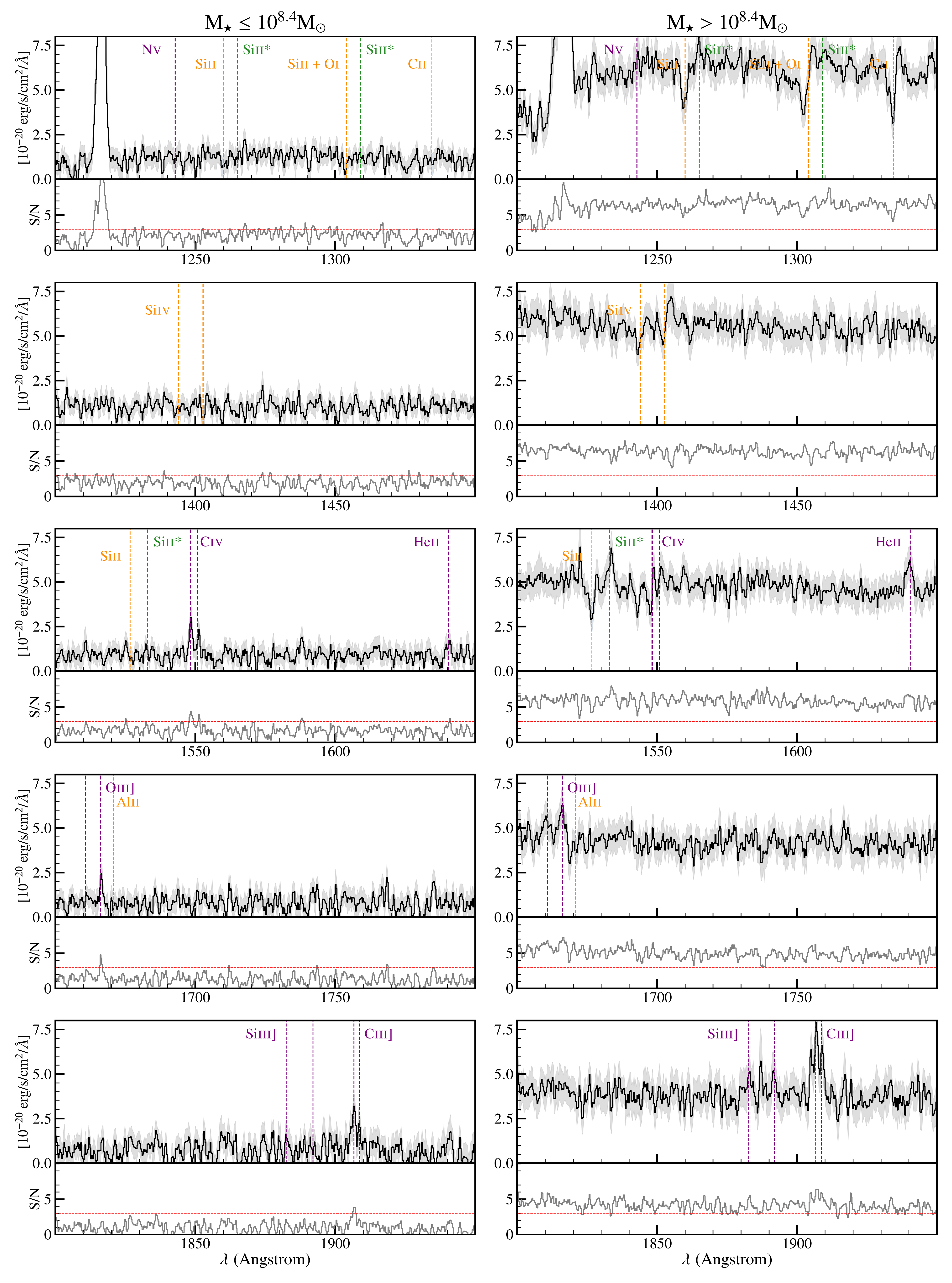}
   \caption{Mean spectra of LAEs with stellar mass $\leq -0.32$ and $> -0.32$ M$_{\odot}$ (left and right, respectively).  The median values of the stellar mass for the two subsamples are $10^{7.9}$ and $10^{9}$ M$_{\odot}$, respectively.   Lines and symbols as in Fig. \ref{Fig3}.}
            \label{FigA7}%
  \end{figure*}

% Figure A8
  \begin{figure*}
 \centering
   \includegraphics[width=16.5cm]{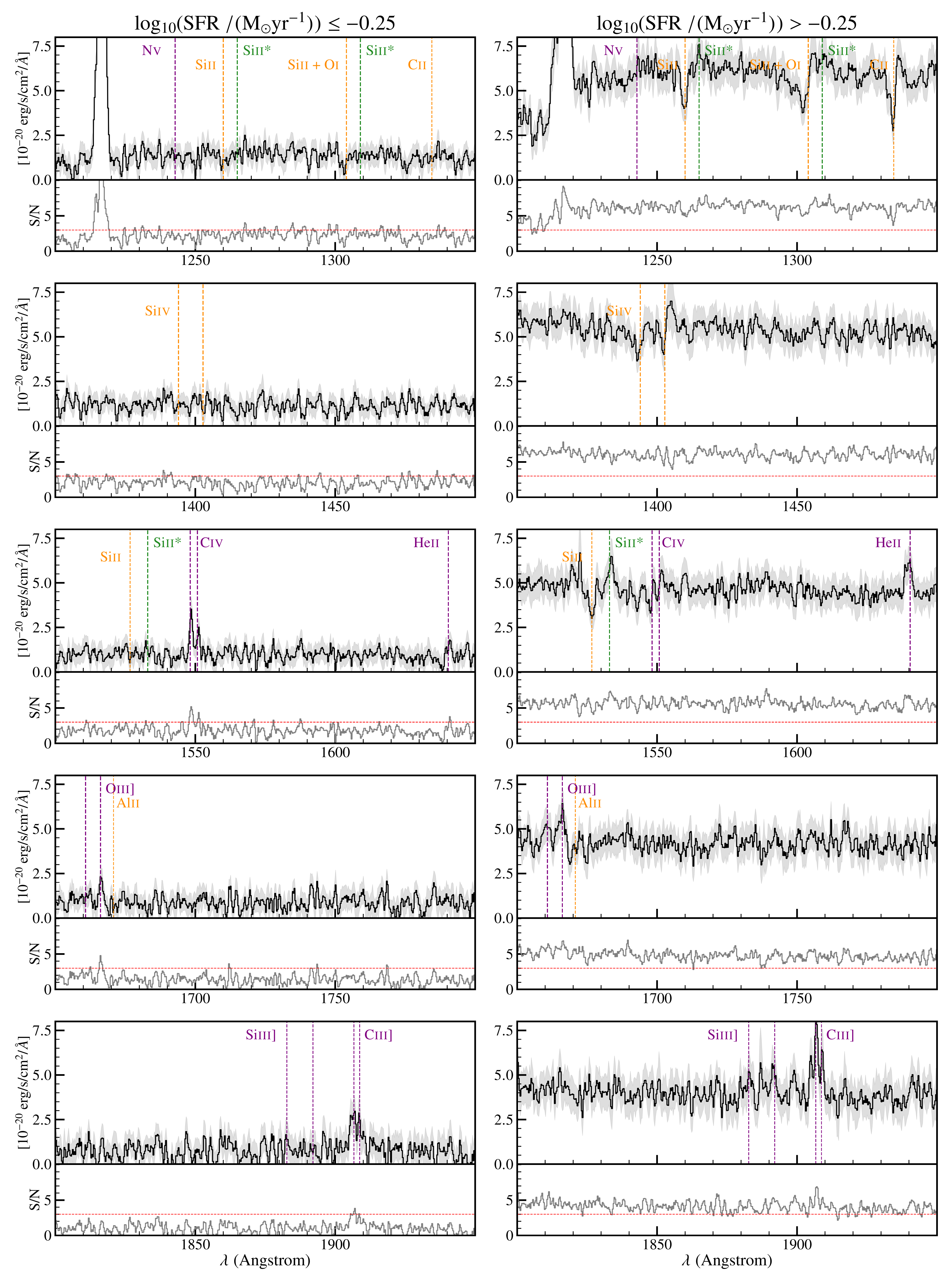}
   \caption{Mean spectra of LAEs with log$_{10}$(SFR / (M$_{\odot}$ yr$^{-1}$)) $\leq -0.25$ and $> -0.25$ (left and right, respectively). The median values of log$_{10}$(SFR / (M$_{\odot}$ yr$^{-1}$)) for the two subsamples are $-0.57$ and $0.16$, respectively. Lines and symbols as in Fig. \ref{Fig3}.}
        \label{FigA8}
  \end{figure*}
%
% Figure A9
  \begin{figure*}
  \centering
   \includegraphics[width=16.5cm]{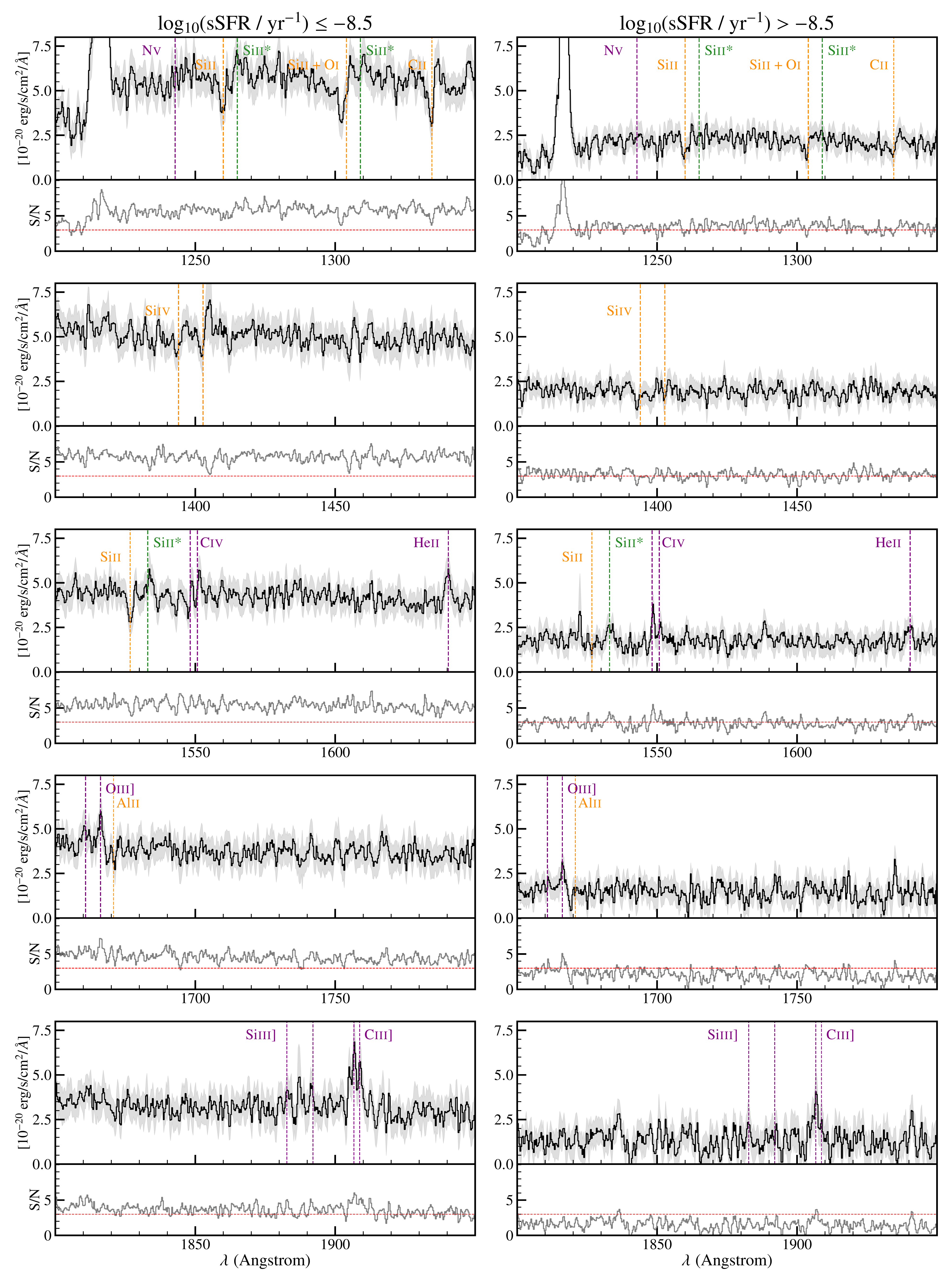}
   \caption{Mean spectra of LAEs with log$_{10}$(sSFR / yr$^{-1}$) $\leq -8.5$ and $> -8.5$ (left and right, respectively). The median values of log$_{10}$(sSFR / yr$^{-1}$) for the two subsamples are $-8.8$ and $-8.5$, respectively. Lines and symbols as in Fig. \ref{Fig3}.}
            \label{FigA9}%
   \end{figure*}

\end{appendix}

%-------------------------------------------------------------------

\end{document}